\documentclass[11pt,a4paper]{article}
\usepackage{jheppub,amsmath,  amssymb,slashed,url,bm,textgreek,upgreek}
\usepackage{graphicx}
\usepackage{epstopdf}
\def\t{{ \sf t}} 

\def\tt{{\mathfrak t}}

\def\zzeta{{\bm \zeta}}

\def\be{\begin{equation}}
\def\ee{\end{equation}}

\def\tilde{\widetilde}

\def\h{\widehat}

\def\red{{\mathrm{red}}}

\def\d{{\mathrm d}}

\def\R{{\mathbb R}}
\def\C{{\mathbb C}}

\def\[{\bigl [}

\def\]{\bigr ]}

\def\Z{{\mathbb Z}}

\def\t{\widetilde }
\def\h{\widehat}

\def\P{{\mathcal P}}

\def\H{{\mathcal H}}

\def\tilde{\widetilde}
\def\i{{\mathrm i}}
\def\la{\langle}
\def\ra{\rangle}

\def\bar{\overline}

\def\E{{\mathbb E}}

\def\max{{\mathrm{max}}}
\def\zzeta{\zeta}

\newcommand{\normord}[1]{:\mathrel{#1}:}

\title{Liouville Theory: 
An Introduction to  Rigorous Approaches}

\author{Sourav Chatterjee$^1$ and Edward Witten$^2$}
\affiliation{$^1$Department of Statistics, Stanford University, Stanford CA 94305 USA\\
$^2$School of Natural Sciences, Institute for Advanced Study,  Princeton, NJ 08540 USA}

\abstract{In recent years, a surprisingly direct and simple rigorous understanding of quantum Liouville theory has developed.   We aim here to make
this material more accessible to physicists working on quantum field theory.  }

\begin{document}\maketitle

\section{Introduction}\label{intro}

Since it was first proposed more than forty years ago, quantum Liouville theory, or just Liouville theory for short,  has had many applications and has
been studied from many points of view.   Relevant papers and review articles  include 
\cite{Polyakov,CT,GN,KPZ,GL,Seiberg,DO,ZZ,Teschner,Teschner2, Nakayama}, along with many others.   Rigorous mathematical work relevant to Liouville theory also has a long history,
and again it is not practical to give full references.    Hoegh-Krohn rigorously constructed a two-dimensional theory of a massive scalar
field with an exponential interaction, using the positivity of a normal-ordered real exponential \cite{HK}.  Kahane \cite{Kahane}, developing ideas of Mandelbrot \cite{Mandelbrot},
gave what can be interpreted as the first rigorous statistical analysis of the  Liouville measure; the motivation for that work was  the statistical  theory of turbulence rather than  relativistic field theory.   Duplantier
and Sheffield \cite{DS} analyzed Liouville theory on a disc and  demonstrated KPZ scaling from a rigorous point of view.   Closer to our focus in the present article,
 David, Kupianen, Rhodes, and Vargas  \cite{DKRV} gave a rigorous definition of Liouville correlation functions on a sphere, 
and this was further developed to a proof of the DOZZ
formula for the Liouville three-point function by Kupianen, Rhodes, and Vargas  \cite{KRV2}, providing a mathematical framework for ideas of Teschner \cite{Teschner}.  
See  \cite{Vargas,BerestyckiPowell,RV,RRV,DMS,Kup} for further results and background  and connections to some of the many other areas of rigorous 
mathematical physics that are related to Liouville theory but not
discussed in the present article.

 The present article
has both a general and a specific purpose.   The general purpose is to provide for physicists a gentle introduction to rigorous arguments concerning  Liouville theory.   The more specific purpose
is to explain how this approach clarifies some otherwise obscure properties of Liouville correlation functions.

One preliminary point  is  that the Liouville action 
\be\label{laction}I=\int_\Sigma \d^2 x\sqrt g\left( \frac{1}{4\pi}\partial_a \phi \partial^a\phi+\frac{Q}{4\pi}R\phi +\mu e^{2b \phi}\right) ,\ee
is real-valued.   Here $\Sigma$ is a closed two-manifold  with local coordinates $x^a,\,a=1,2$,  
metric $g$ and scalar curvature $R$;  $b>0$ is the Liouville coupling,\footnote{\label{conventions}We follow conventions of \cite{ZZ,Nakayama}, for example, but we should note that much of the rigorous literature is expressed in terms of $\gamma =2b$.  Similarly, as in \cite{ZZ,Nakayama} and much
recent literature, we denote Liouville primary fields as $e^{2\alpha\phi}$; much of the rigorous literature is expressed in terms of $\upalpha=2\alpha$.}  and $Q=b+1/b$; and $\mu>0$ is the ``cosmological constant,'' whose
value is inessential as it can be adjusted by KPZ scaling \cite{KPZ}.   Because $I$ is real, the path
integral measure $D \phi \,e^{-I}$ is positive-definite, and probabilistic methods can potentially be applied.    
However, this is hardly special to Liouville theory.   Many other
theories, such as $\phi^4$ theory in any spacetime dimension, or four-dimensional gauge theory with vanishing theta-angle, likewise have a real action and a positive path integral
measure.

Much more special to Liouville theory is that the interaction is positive, even quantum mechanically. Classically, the function $e^{2 b\phi}$, with real $b$ and $\phi$, is positive, of course. 
Quantum mechanically, in the case of Liouville theory, the only renormalization that this interaction requires  is normal-ordering with respect to an underlying free field theory measure.
Normal-ordering of an exponential interaction such as $e^{2b\phi}$ gives a multiplicative renormalization, which preserves positivity.   So the renormalized interaction is positive,
and this plays a crucial role.

To appreciate how exceptional is the positivity of the Liouville interaction, let us consider some other possible perturbations of a free massless scalar field $\phi$.
A mass term $\frac{1}{2}m^2\phi^2$ can again be renormalized by normal-ordering.   In this case, normal-ordering entails the subtraction of a divergent constant -- an additive
rather than multiplicative renormalization.   After this subtraction, $\frac{1}{2}m^2\phi^2$ is a well-defined observable,  but unbounded below: one cannot make it positive
definite by adding a constant.  Similarly
any polynomial function $P(\phi)$ of degree greater than 1 is not bounded below after normal-ordering.   The exponential  interaction in Liouville theory is special.\footnote{Other
theories of scalar fields in two dimensions with a potential that is  a sum of real exponentials, such as sinh-Gordon theory with
potential $V(\phi)=\mu \cosh 2b\phi$, have the same positivity. This class of theories was studied in \cite{HK}.}  

In Liouville theory, one usually wants to calculate the correlation function of a product of primary fields
$e^{2\alpha_i\phi(x_i)}$, $i=1,\dots,n$.   For our purposes in this article, the parameters $\alpha_i$ are real, to ensure that the primary fields $e^{2\alpha_i\phi(x_i)}$ are positive,
as assumed in the probabilistic approach.  But we do not assume that the $\alpha_i$ are positive.   We will eventually impose the Seiberg
bound $\alpha_i\leq Q/2$,  after understanding why it is needed.  
We will concentrate on correlation functions for the case that $\Sigma$ has genus $0$.   As the genus $0$ partition function does not converge, it is usual to define unnormalized
correlation functions, without trying to divide by the partition function:
\be\label{unnormalized}\left\la \prod_{i=1}^n e^{2\alpha_i\phi(x_i)}\right\ra=\int D\phi \,e^{-I(\phi)} \,   \prod_{i=1}^n e^{2\alpha_i\phi(x_i)}.\ee
Let us recall a few relevant facts.  
Goulian and Li \cite{GL} showed that  the path integral in eqn.~(\ref{unnormalized}) 
 can be usefully studied by first integrating over the ``zero-mode'' of the Liouville field.   
 One writes
 \be\label{splitting}\phi=c+X, \ee
 where $c$ is a constant, and $X$ is a real-valued field that is subject to one real constraint (which can be chosen in various ways).   The integral over $c$
 can be done explicitly,  reducing the evaluation of Liouville correlation functions to an integral over $X$.  In genus $0$, after changing variables from $X$ and $c$ to $X$ and 
  $t=e^{2bc}\mu\int \d^2x\sqrt g e^{2bX}$ and integrating over $t$, and using the fact that $\int_\Sigma\d^2x\sqrt g  R=8\pi$, one gets
 \be\label{redform}\left\la \prod_{i=1}^n e^{2\alpha_i\phi(x_i)}\right\ra =\frac{\Gamma(-w)}{2b}\int D X \,e^{-I_\red(X)}\prod_{i=1}^n e^{2\alpha_i X(x_i)} \left(\mu\int\d^2x\sqrt g e^{2bX(x)}\right)^w, \ee
 where 
 \be\label{mmtdegree} w=\frac{Q-\sum_i\alpha_i}{b} \ee
 and the reduced action for $X$ is a free field action
 \be\label{redaction}I_\red(X)=\frac{1}{4\pi}\int\d^2x\sqrt g   \left( \partial_a  X\partial^a X+ QRX\right). \ee
 Thus ``all'' we have to do to determine  Liouville correlation functions is to compute the expectation value of $\prod_{i=1}^n e^{2\alpha_i X(x_i)}
 \left(\mu\int\d^2x\sqrt g e^{2bX(x)}\right)^w$ with respect to the
 free field measure $D X\,e^{-I_\red(X)}$.   The only problem is that standard methods suffice to understand and potentially to compute this expectation value only when $w$ is
 a non-negative integer.
 
 When $w$ is a non-negative integer, the gamma function in eqn.~(\ref{redform}) has a pole.   We will call these poles ``perturbative poles,'' as they can be understood
 by the simple semiclassical calculation that we have just explained, or alternatively since they can be understood in a perturbation expansion in powers of $\mu$. 
Goulian and Li were able to compute the residue of the correlation function at  a semiclassical pole, by using the fact that in this case the exponent $w$ is a non-negative integer, and
making use of integrals that had been computed earlier \cite{DF}.

  The Liouville three-point function in genus $0$ is particularly important as it is expected to determine the whole theory by bootstrap or factorization arguments.
By looking for a relatively simple formula
that incorporates the perturbative poles and residues, and also satisfies recursion relations that those poles and residues obey, 
Dorn and Otto \cite{DO} and A. and Al. Zamolodchikov \cite{ZZ}  suggested  an exact formula (the DOZZ formula) for this three-point function.  
The DOZZ formula, which will be described in somewhat 
more detail in section \ref{dozzcompar}, was put on a much more solid basis by Teschner \cite{Teschner}, helping provide the basis for a  
rigorous proof  of this formula \cite{KRV2}.  

There was  guesswork in the original discovery of the DOZZ formula, and this discovery also created  a puzzle:
although the DOZZ formula for the Liouville three-point function was motivated by trying to reproduce the perturbative poles and their residues, 
it actually has many additional poles whose interpretation
has been less obvious.   
We may call the poles  that appear in the DOZZ formula and do not simply arise from the integral over the Liouville zero-mode ``nonperturbative,'' as they cannot be understood 
in a simple perturbation expansion in powers of $\mu$ (or $b$).    
A qualitative or semiclassical understanding of the nonperturbative poles has been missing.

In this article,  we will explain how the rigorous probabilistic approach to Liouville theory sheds some light on those nonperturbative poles.   A detailed explanation, which will be given
in the body of this article, requires
understanding non-integer powers of the Liouville interaction $\mu \int_\Sigma \d^2x\sqrt g  \,e^{2b\phi}$, and related functions that appear in the analysis.  
In this introduction, however, we will give a heuristic explanation, in which we consider only integer powers.  

Are there ultraviolet divergences in Liouville theory?  
Consider expanding Liouville correlation functions  in powers of $\mu$.   In $n^{th}$ order, we have to consider an $n$-fold integral
of the  normal-ordered interaction $\normord{e^{2 b \phi}}$, of the general form $\int \d^2 y_1\cdots \d^2 y_n\normord{e^{2b\phi(y_1)}}\cdots \normord{e^{2b\phi(y_n)}}$.    
Let us first consider a possible ultraviolet divergence that occurs when the points $y_1,\ldots,y_n$ over which we integrate coincide at a point $y_0\in \Sigma$ 
that is disjoint from all the points $x_i$ at which external vertex operators $\normord{e^{2\alpha_i\phi(x_i)}}$ are inserted.   
In looking for such  a  divergence, we can restrict the $y_i$ to a compact region $A\subset \Sigma$ that does not contain any of the $x_i$
and thus we consider the integral
\be\label{nfold} \int_{A^n}  \d^2 y_1\cdots \d^2y_n  \, \normord{e^{2b\phi(y_1)}}\,\normord{e^{2b\phi(y_2)}}\cdots \normord{e^{2b\phi(y_n)}} .\ee
At short distances, $\phi$ can be viewed as  a free field with two-point function $\la\phi(x)\phi(y)\ra=\log \frac{1}{|x-y|}$, and the expectation value of this  integral becomes \be\label{nfold2}\int_{A^n}\d^2 y_1\cdots \d^2 y_n \prod_{1\leq i<j\leq n}\frac{1}{|y_i-y_j|^{4b^2}}. \ee
Along the diagonal $y_1=y_2=\cdots =y_n$, the integral scales as $\int_{|y|\leq 1}\d^{2n-2} y\, |y|^{-2n(n-1)b^2}$, and diverges if and only if $n\geq 1/b^2$.  (The behavior along subdiagonals
where only some of the $y'$s coincide is less singular.)    From the point of view of the operator product expansion, this divergence  can be understood as follows.
On general
grounds, one expects such an integral  to diverge if and only if a relevant or marginal operator appears in the product of $n$ copies of $\normord{e^{2b\phi}}$.   The lowest dimension
operator that appears in that $n$-fold product is $\normord{e^{2bn\phi}}$.   Recalling that in Liouville theory, the (holomorphic or antiholomorphic) dimension 
of the operator $\normord{e^{2\alpha\phi}}$ 
is $\h\Delta_\alpha=\alpha(Q-\alpha)$, we see that the condition in Liouville theory to get an ultraviolet divergence in $n^{th}$ order of perturbation theory in $\mu$ is 
$nb(Q-nb)\leq 1$, or $n\geq 1/b^2$.   In the body of this article, we will explain how this analysis can be extended rigorously to non-integer values of $n$, and that will account 
for one of the nonperturbative poles of the DOZZ formula, namely the pole at $Q-\sum_i\alpha_i=1/b$ or $w=1/b^2$.  

We can also consider the case that multiple copies of the interaction collide with one of the external vertex operators.   For this, we assume the region $A$ to contain one of the points $x$ 
 at which an external vertex operator $\normord{e^{2\alpha\phi(x)}}$ is inserted.  Now we consider the integral
\be \label{modfold} \notag\int_{A^n} \d^2 y_1 \d^2y_2 \cdots \d^2y_n  \,\normord{e^{2\alpha\phi(x)}} \,\normord{e^{2b\phi(y_1)}}\,\normord{e^{2b\phi(y_2)}}\cdots \normord{e^{2b\phi(y_n)} },\ee
and the expectation value of this expression in the free field approximation is
\be\label{thirdmod} \int_{A^n} \d^2 y_1 \d^2y_2 \cdots \d^2y_n\prod_{k=1}^n\frac{1}{|x-y_k|^{4\alpha b}}\prod_{1\leq i<j\leq n}\frac{1}{|y_i-y_j|^{4b^2}} .\ee   This integral diverges when all $n$ points $y_1,\ldots, y_n$ approach $x$
 if and only if $n>1/b^2+1-2\alpha=(Q-2\alpha)/b$.
From the point of view of the operator product expansion, one expects such a divergence if and only if an operator
whose dimension is no greater than that of $e^{2\alpha\phi}$ appears in the product of $e^{2\alpha\phi}$ with $n$ copies of $e^{2b\phi}$.  The lowest dimension operator that appears in that
product is $e^{2(\alpha+nb)\phi}$, so the condition for a divergence is $\h\Delta_{\alpha+nb}\leq \h\Delta_\alpha$ or $n\geq (Q-2\alpha)/b$.    When properly extended to
non-integer values of $n$, this analysis accounts for another nonperturbative pole of the DOZZ formula.  

This analysis will not explain all of the nonperturbative poles of the DOZZ formula, but only those that occur on the boundary of a certain accessible region.   
Actually  for $\Sigma$ of genus $0$, the path integral  (\ref{unnormalized})
that defines  the Liouville correlation functions converges only if $w<0$ or $\sum_i\alpha_i>Q$.  Otherwise, the integral over $c$ diverges, leading to a pole at $w=0$
and (after analytic continuation past this pole) to the other perturbative  poles at $w=1,2,3,\ldots$.   After integrating over $c$ and extracting the explicit  factor $\Gamma(-w)$ in
eqn.~(\ref{redform}), the remaining integral over $X$  has an extended region of convergence.   
But it is not obvious  what is this extended region, since it is bounded by nonperturbative phenomena.   In the body of this article, we will explain how the extended
region of convergence has been determined rigorously by probabilistic arguments.   Two boundaries of the extended region of convergence are associated with nonperturbative
poles of the DOZZ formula.   The locations of these boundaries can be guessed by assuming that the preceding analysis of ultraviolet divergences  can be  generalized to non-integer
values of $n$.   Perhaps surprisingly, the extended region of convergence has a third boundary  that is associated with a zero rather than a pole of the DOZZ formula.
This zero occurs when one of the $\alpha_i$ reaches the Seiberg bound at $Q/2$.   One might be surprised that an ultraviolet effect can lead to a zero rather than a pole.   What makes this possible is that when the exponent $w$ is negative,
 an ultraviolet effect,  a sort of divergence of the denominator in eqn.~(\ref{redform}), can lead to a vanishing of the correlation function.   It is not clear to us whether based on 
heuristic arguments one should expect that the DOZZ formula would vanish when one of the $\alpha_i$ reaches the Seiberg bound. But at any rate this does 
happen, and the probabilistic analysis gives a nice explanation of why.

To more fully 
understand  the nonperturbative behavior of the DOZZ formula, one would need to  understand the  analytic continuation of the path integral beyond its extended region of convergence.
Even without that,  the results that we have summarized
may provide an interesting general hint about  quantum gravity.   In recent years, there has been much interest in possible unexpected  
breakdown of semiclassical behavior of gravity in complex
situations with many particles.   Bearing in mind that small $b$ is the semiclassical regime of Liouville theory,  ultraviolet divergences that begin at roughly order $1/b^2$  in
perturbation theory may be a somewhat similar phenomenon.

In section \ref{correlators} of this article, we give an overview of aspects of Liouville theory correlators from a probabilistic point of view.   In section \ref{detailedarguments},
we explain more detailed arguments.

\section{Liouville Theory Correlators: Overview}\label{correlators}

\subsection{The Gaussian Free Field}

Though Liouville theory is conformally invariant,  to analyze  Liouville theory correlators in genus $0$, that is for the case 
that the two-manifold $\Sigma$ is a sphere, it is convenient to pick a specific Kahler metric on $\Sigma$.   An obvious choice would be a ``round'' metric, but it has been found
that it is convenient to pick a metric in which $\Sigma$ is built by gluing together two unit disks, each with a flat metric, along their common boundary.    To describe such a metric
explicitly, we can view $\Sigma$ as the complex $x$-plane, plus a point at infinity, with the metric
\be\label{kmetric}\d s^2=\d x\d\bar x \cdot\begin{cases} 1 &  |x|\leq 1\cr \frac{1}{|x|^4}&|x|>1.\end{cases}\ee
The region $|x|\leq 1$ is obviously  a unit disc with a flat metric, and as the given metric is invariant under the inversion $x_i\to x_i/|x|^2$, the region $|x|\geq 1$ is another copy of the same thing.    The scalar curvature $R$ of this metric has delta function support on the unit circle $|x|=1$, and is  invariant under rotation of the circle. In polar coordinates $x=r e^{\i\theta}$,
\be\label{rsupport}R=4\delta(r-1).\ee 
It is also useful to make a specific choice of the constraint on $X$ in the decomposition of the Liouville field as $\phi=c+X$, where $c$ is a constant and $X$ must satisfy one real constraint
to make the decomposition unique.   One obvious choice would be to say that the average value of $X$ on the sphere should vanish.   Another obvious choice would involve a constraint
on the ``area''  $\int_\Sigma \d^2 x\sqrt g\,e^{2b\phi(x)}$.     A rather less obvious choice is to ask that
the average of $X$ on the unit circle $|x|=1$ should vanish: 
\be\label{condx} \int_0^{2\pi}\d\theta \,X(e^{\i\theta})=0. \ee
It turns out that this choice simplifies the analysis of the correlation functions.   One preliminary reason for this is that, combining (\ref{rsupport}) and (\ref{condx}), we have 
$\int_\Sigma \d^2 x\sqrt g  R X =0$, and therefore the reduced action (\ref{redaction}) becomes purely quadratic
\be\label{quadact}I_\red=\frac{1}{4\pi}\int\d^2x \sqrt g \partial_a X\partial^a X, \ee
so that $X$ is a Gaussian free field, with mean zero.

This also happens if we choose a round metric on the sphere and constrain $X$ so that its average on the sphere vanishes.   However, the choice involving a circle
average also leads to a useful simplification of the propagator or two-point function  $G(x,y)= \la X(x) X(y)\ra $, which is
\be\label{twopoint}  G(x,y)=\log \frac{1}{|x-y|} +\log |x|_+ +\log |y|_+,\ee
where $|x|_+=\max(1,|x|)$.  In verifying this, and in later applications, it is useful to know the following
formula  for an integral over a circle of radius $r$ centered at a point $x\in \C$:
\be\label{usefulfact}\frac{1}{2\pi}\int_0^{2\pi}\d\theta \log|x+r e^{\i\theta}-y| =\begin{cases} \log |x-y| & |x-y|\geq r\cr
                                                                                                                                              \log r & |x-y|\leq r.\end{cases}\ee
 With $\log |x-y|$ as the two-dimensional analog of the Newtonian gravitational potential, this formula is the two-dimensional analog of Newton's discovery that (in modern language)
 the gravitational potential of a spherical shell is that of  a point mass outside the shell, and is constant inside the shell. 
 With the help of eqn.~(\ref{usefulfact}), one can see that eqn.~(\ref{twopoint}) is consistent with the constraint (\ref{condx}), since it implies that $ \int_0^{2\pi}\d\theta \la
 X(e^{\i\theta})X(y)\ra=\int_0^{2\pi}\d \theta G(e^{\i\theta},y)=0$ for all $y$.
 The formula (\ref{twopoint}) follows from this along with the fact that $G(x,y)$ has the expected logarithmic singularity along the diagonal at $x=y$, and satisfies the appropriate
 differential equation $-\nabla_x^2 G(x,y)=2\pi\delta^2(x-y)+\lambda \delta_{|x|=1}$ (where $\lambda$
is a Lagrange multiplier related to the constraint at $|x|=1$).

Of course, placing the constraint on the circle $|x|=1$ is  arbitrary.   For example, to place the constraint instead on a circle $|x|=T$, we would replace
$X$ with a new Gaussian free field
\be\label{newgff}\t X(x)=X(x) -\frac{1}{2\pi}\int_0^{2\pi}\d \theta X(T e^{\i\theta}), \ee
now satisfying
\be\label{newconstraint}\int_{0}^{2\pi} \d\theta\t X(T e^{\i\theta})=0 . \ee
The propagator or two-point function  $\la\t X(x)\t X(y)\ra=\t G(x,y)$ is given by
\be\label{newtwo} \t G(x,y)=\log \frac{1}{|x-y|} +\log \max(T,|x|)+\log\max(T,|y|)-\log T. \ee

  \subsection{Preliminary Steps}                                                                                                                          

Our goal  is to show that the formula (\ref{redform}) for the Liouville correlation functions makes sense (after imposing some restrictions on the $\alpha_i$).  
A useful preliminary reduction, which mathematically can be viewed as an application of Girsanov's theorem, is the following.   We can eliminate the factors
$\prod_{i=1}^n e^{2\alpha_i X(x_i)} $ in eqn.~(\ref{redform}) by ``completing the square'' in the product
\be\label{actprod}e^{-I_\red(X)} \prod_{i=1}^n e^{2\alpha_i X(x_i)} =\exp\left(-\frac{1}{4\pi}\int \d^2 x \sqrt g  \partial_a X\partial^a X\right) \prod_{i=1}^n e^{2\alpha_i X(x_i)}.\ee
This is done by shifting $X(x)\to X(x)+2\sum_i\alpha_i G(x,x_i)$, leading to\footnote{The operators $e^{2\alpha_i X(x_i)}$ need to be normal-ordered, as we will do in a moment for $e^{2b X}$.   As usual, taking this normal ordering into account eliminates divergent factors involving $G(x_i,x_i)$ that would otherwise appear in the following formula.}
\be\label{newform} 
\begin{split} &\left\la \prod_{i=1}^n e^{2\alpha_i\phi(x_i)}\right\ra \\
&=\frac{\Gamma(-w)\mu^w}{2b}\int DX \,e^{-I_\red(X)}\prod_{i<j}\left(\frac{|x_i|_+|x_j|_+}{|x_i-x_j|}  \right)^{4\alpha_i\alpha_j}\left( \int_\Sigma \d^2x \sqrt g  f(x) e^{2bX(x)}\right)^w,
\end{split} \ee   
where
\be\label{multfn}f(x) =\prod_{i=1}^n \left(\frac{|x|_+|x_i|_+}{|x-x_i|}\right)^{4b\alpha_i }.\ee
In other words, the Liouville correlator, up to some explicitly known factors, can be defined by the $w^{th}$ moment of the observable
\be\label{intobs} M_f =\int_\Sigma \d^2x \sqrt g  \,f(x) e^{2bX(x)} \ee
with respect to the free field measure.   
More generally, we will consider the restriction of such an integral to a region $A\subset \Sigma$:
\be\label{resta}M_f(A)=\int_A\d^2x \sqrt g \,f(x) e^{2bX(x)}.\ee
If $w$ is a non-negative  integer, then the $w^{th}$ moment of $M_f(A)$ is simply the expectation value in the free field measure of a $w$-fold integral of $f(x) e^{2b X(x)}$ over
$A$.   This can be evaluated, or at least reduced to an explicitly defined integral, by using the two-point function (\ref{twopoint}) to  evaluate the expectation value 
of a product of exponentials. That is what Goulian and Li did \cite{GL}, with a different choice of the Kahler metric and the condition on $X$.
  What can we do when $w$ is not a non-negative integer?   This is where probabilistic methods \cite{DKRV,KRV,KRV2,Vargas,BerestyckiPowell} have been useful,
as we will aim to explain.

In studying $M_f(A)$ as a random variable, it turns out  that if the  Seiberg bound $\alpha_i\leq Q/2$ is not satisfied, then $M_f(A)=+\infty$ with probability 1.  
In that case, probabilistic methods (or at least the arguments that we will explain) do not enable one to make sense of Liouville correlation functions.
However, as long as   all $\alpha_i$ satisfy the Seiberg bound, it is possible to define the $M_f$ (or  $M_f(A)$ for any $A$) as a positive random variable.  
 Saying that $M_f$ is  a positive random variable means that there is a probability  measure $\rho$ on the positive real line that determines the expectation value of a function of $M_f$.  
 By definition a probability measure on the half-line is non-negative and satisfies 
\be\label{probmeas}\int_0^\infty \d u \,\rho(u)=1.\ee
The function $\rho$ is not necessarily smooth; it might have delta function support at some values of $u$. In Liouville theory, there can be a delta function at $u=0$ (but not elsewhere). 
This happens  precisely if
$b>1$, in which case it turns out that $M_f=0$ with probability 1, corresponding to  $\rho(u)=\delta(u)$.   So for $b>1$ the methods we will describe will  again not  be useful
for defining Liouville correlation functions.  
 
In general, the expectation value
of a function $h(\Phi)$ of a random variable $\Phi$ with probability measure $\rho$ is 
\be\label{expval} \E[h(\Phi)] =\int_0^\infty \d u \rho(u) h(u), \ee
assuming that this integral is well-defined.    
Here we switch to probabilistic notation and write $\E[h(\Phi)]$ rather than $\la h(\Phi)\ra$ for the expectation value of $h(\Phi)$.   
If the function $h$ is not positive or negative-definite, the integral in eqn.~(\ref{expval}) may be ill-defined.   However, in our application, $h$ will be a positive function
$h(u)=u^w$ for some real $w$.   Given this, assuming that $w>0$ or that $\rho(u)$ does not have a delta function at $u=0$ (which would make the integral ill-defined if $w<0$), the integral for $\E[M_f^w]$ converges
to a positive real number or $+\infty$.   When the integral converges to $+\infty$, this will mark a boundary of the extended region of convergence that was discussed
in the introduction: on this boundary, the DOZZ formula will have a nonperturbative singularity.   If $\E[M_f^w]$ vanishes for some $w>0$, this will tell us that $\rho(u)=\delta(u)$
and that $M_f$ vanishes with probability 1.   This happens  if $b>1$.   Finally, $\E[M_f^w]$ will converge for negative $w$ if and only if $\rho(u)$ vanishes sufficiently 
rapidly near $u=0$.  In Liouville theory, the random variable $M_f$ always has that property.

It will be convenient if we can take the two-point function of $X$ to be simply $\log(1/|x-y|)$, not just because this is a simple function but more importantly
because it has nice scale invariance properties and a useful relation to random walks.   To reduce to this simple form of the propagator, we can note the following.   A finite sum of positive random variables
is again a positive random variable.   To exploit this, we can decompose $\Sigma$ as a union of 
two sets $\Sigma_1$ and $\Sigma_2$, where $\Sigma_1$ is the unit disc $|x|\leq 1$, and $\Sigma_2$ is the set $|x|\geq 1$.
Then $A=A\cap \Sigma_1\cup A\cap \Sigma_2$, so
\be\label{randomsum}  M_f(A) =M_f(A\cap \Sigma_1)+M_f(A\cap \Sigma_2).\ee
Hence defining the $M_f(A\cap \Sigma_i)$ as positive random variables for $i=1,2$ will lead to such a definition for $M_f(A)$.
Eqn.~(\ref{twopoint}) already has the desired form for $x,y\in A\cap \Sigma_1$, and the same is true in $A\cap \Sigma_2$ after a change
of variables $x\to 1/x$.  Another convenient fact is that in each of the two regions $\Sigma_1$ and $\Sigma_2$, the metric is simply the Euclidean metric $\d x_1^2+\d x_2^2$
(up to the inversion $x_i\to x_i/|x|^2$ in the case of $\Sigma_2$).

 Of course, there is nothing special about the particular decomposition $\Sigma=\Sigma_1\cup\Sigma_2$ that was used here.   
We could have used the variable $\t X$ instead
of $X$, making a similar division at $|x|=T$, and we could  similarly decompose $\Sigma$ as the union of more than two pieces. 

So finally, to define Liouville correlators, it suffices to consider a region  $A\subset \Sigma$ in which  the metric is flat
and the two-point function of $X$ is a simple logarithm, 
and to study the quantity
\be\label{studyfun} M_f(A)=\int_A \d^2x    f(x) e^{2b X(x)}, \ee
where the function $f(x)$ is either  smooth and bounded  in region $A$, 
or has just one singularity of the form $|x-x_0|^{-4b\alpha}$ for some $\alpha$ and some $x_0\in A$.

\subsection{Normal Ordered Interaction}\label{normalordered}

The field $X(x)$ can be understood as a random Gaussian distribution.   This statement means that if we pair it with a smooth function $f(x)$ (real-valued for simplicity,
and necessarily of compact support
as $\Sigma$ is compact), 
we get a Gaussian random variable $X_f=\int\d^2x  f(x) X(x)$, with $\E[X_f]=0$ and $\E[X_f^2]=\int\d^2x  \d^2y    f(x)f(y) G(x,y)$.
This integral converges for any $f$.  Without such smearing,
although the ``operator'' $X(x)$
has well-defined correlation functions $\E[X(x)X(y)]=G(x,y)$ for $x\ne y$, it does not make sense as a random variable.  Its variance would be $\E[X(x)^2]=G(x,x)=+\infty$, but a Gaussian measure
$\rho(u)=\frac{1}{\sigma \sqrt{2\pi}} e^{-u^2/2\sigma^2}$ does not have a limit as the variance $\sigma^2$ goes to infinity.  So one cannot assign probabilities to the values of $X(x)$;
it does not really have values.

The random distribution that we want to define is not $X(x)$ but a renormalized version of $e^{2b X(x)}$.  Here, renormalization is required just to define correlation functions of
$e^{2b X(x)}$,  even without trying to define it as a random variable.   To accomplish this, we will first introduce a cutoff by smearing $X$ slightly to get a cutoff version $X_\epsilon$ that is a Gaussian
random variable, so that $e^{2 b X_\epsilon(x)}$ will also make sense as a random variable.   Then we will renormalize by normal-ordering.  We will also describe another
useful kind of cutoff in section \ref{martingale}.

 In general, if $Y$ is a centered Gaussian
variable (the statement that it is ``centered'' means that its expected value is zero), the normal ordered version of $Y$ is $e^{Y-\frac{1}{2}\E[Y^2]}$, 
which we will denote as $\normord{e^Y}$.   Note that  $\E[\normord{e^Y}]=1$ for any $Y$.
 In particular,  the normal-ordered version of $e^{2b X_\epsilon(x)}$ will be
$\normord{e^{2b X_\epsilon(x)}}=e^{2b X_\epsilon(x)-2b^2\E[X_\epsilon^2]}$.  Finally we will define
\be\label{correcteddef} M_{f,\epsilon}(A)=\int_A\d^2x  f(x) \normord{e^{2bX_\epsilon(x)}}\ee
and investigate the behavior as  $\epsilon\to 0$.  The random variable that we want is
\be\label{limitdef}M_f(A)=\lim_{\epsilon\to 0} M_{f,\epsilon}(A),\ee
if this limit exists.   If the limit does exist, then it makes sense to define the moments $E[M_f(A)^w]$ of $M_f(A)$  with $w$ not necessarily an integer, 
though in general these moments may equal $+\infty$.   Similar we can take a linear combination of random variables $M_f(A)$ with positive coefficients
and define their moments for general real $w$.

One obvious way to define a smeared version $X_\epsilon(x)$ of $X(x)$  would be to smear $X(x)$ over a disc of radius $\epsilon$ centered at $x$.   However, it turns out that
it is more convenient to instead average $X(x)$ over a circle of radius $\epsilon$ centered at $x$.  Thus we define
\be\label{circlereg} X_\epsilon(x) =\frac{1}{2\pi}\int_0^{2 \pi} \d\theta X(x+ \epsilon e^{\i\theta}). \ee
If $\la X(x)X(y)\ra=\log\frac{1}{|x-y|}$ in the relevant portion of $\Sigma$, then for $|x-y|>2\epsilon$, we have simply $\la X_\epsilon(x)X_\epsilon(y)\ra=\log \frac{1}{|x-y|}$,
just as at $\epsilon=0$.  For $|x-y|<2\epsilon$, that is no longer true, since the circles of radius $\epsilon$ centered at $x$ and $y$ intersect.  For example,
\be\label{specialcase} \la X_\epsilon(x) X_\epsilon(x)\ra=\log \frac{1}{\epsilon},\ee
 by use of eqn.~(\ref{usefulfact}).   
 
 The general formula for $\la X_\epsilon(x) X_\epsilon(y)\ra$ is a little complicated, but what we will need to know can be summarized in a few simple facts.   
First, this function  is monotonically increasing as $\epsilon$ becomes smaller.  Equivalently,
\be\label{monotonic}\frac{\partial}{\partial\epsilon}\la X_\epsilon(x) X_\epsilon(y)\ra  \leq 0\ee
for all $x,y$.  Indeed, more generally, for any $\epsilon,\epsilon'>0$, 
\be\label{moregenfact}\frac{\partial}{\partial \epsilon} \la X_\epsilon(x) X_{\epsilon'}(y)\ra \leq 0,\ee  so $ \la X_\epsilon(x) X_{\epsilon'}(y)\ra$ increases as either
$\epsilon$ or $\epsilon'$ becomes smaller.   To show that, first note from eqn.~(\ref{usefulfact}) that
\be\label{onotone} \la X_\epsilon(x) X(y)\ra =\min\left(\log\frac{1}{|x-y|},\log \frac{1}{\epsilon}\right),\ee
and clearly this is monotonically increasing as $\epsilon$ decreases.   Therefore, averaging over $\theta$,
$\la X_\epsilon(x) X_{\epsilon'}(y)\ra=\frac{1}{2\pi}\int_0^{2\pi}\d\theta\left \la X_\epsilon(x) X(y+\epsilon' e^{\i\theta})\right\ra $ is monotonically increasing
as $\epsilon$ decreases.   By symmetry, the same is true as $\epsilon'$ decreases.  Setting $\epsilon=\epsilon'$, we deduce the inequality~(\ref{monotonic}), which will be used in section \ref{detailedarguments}.
A second useful fact is 
\be\label{secondineq}\la X_\epsilon(x) X_\epsilon(y)\ra\leq \min\left(\log\frac{1}{|x-y|},\log \frac{1}{\epsilon}\right) \ee
(with equality if and only if $|x-y|\geq 2\epsilon$ or $x=y$).   This statement is equivalent to two inequalities: the left hand side is equal to or less than
$\log \frac{1}{|x-y|}$ and equal to or less than $\log\frac{1}{\epsilon}$.  The first inequality is true because it is true (as an equality) if $\epsilon=0$, and therefore
by eqn.~(\ref{monotonic}), it is also true for $\epsilon>0$.   Concerning the  second inequality, by symmetry we can assume $x=0$, $y=r$.
By eqn.~(\ref{specialcase}),  the second inequality holds as an equality if $r=0$, so it suffices to show that
\be\label{lastcase}\frac{\partial}{\partial r}\la X_\epsilon(0)X_\epsilon(r)\ra \leq 0.\ee
We have 
\be\label{dblintegral} \la X_\epsilon(0)X_\epsilon(r)\ra   = \frac{1}{4\pi^2}\int_0^{2\pi}\d\theta\int_0^{2\pi}\d\theta' \la X(\epsilon e^{\i \theta} )X(r+\epsilon e^{\i\theta'})\ra.\ee
Integrating over $\theta$ via  eqn.~(\ref{usefulfact}),  we get $\la X_\epsilon(0)X_\epsilon(r)\ra= \frac{1}{2\pi}\int_0^{2\pi}\d\theta' w(r,\theta')$, where
\be\label{twochoices} w(r',\theta)=\min\left(\log\frac{1}{|r+\epsilon e^{\i\theta'}|}, \log \frac{1}{\epsilon}\right).\ee
Then $\frac{\partial w(r,\theta')}{\partial r}\leq 0$ for all $r,\theta'$, because $\frac{\partial}{\partial r} |r+\epsilon e^{\i\theta'}|$ is positive except when 
$|r+\epsilon e^{\i\theta'}|<\epsilon$, in which case  $w=\log(1/\epsilon)$ and $\frac{\partial w}{\partial r}=0$.   Averaging the result $\frac{\partial w}{\partial r}\leq 0$ over $\theta'$,
we get the claimed inequality~(\ref{lastcase}).  Eqn. (\ref{secondineq}) can be generalized to an upper bound on $\la X_\epsilon(x)X_{\epsilon'}(y)\ra$.   The case of this that we will
need is
\be\label{uselater}\la X_\epsilon(x) X_{\epsilon'}(y)\ra =\log \frac{1}{|x-y|} ~~{\mathrm {if}}~~ |x-y|\geq 2\epsilon, 2\epsilon'.\ee
This follows from eqn. (\ref{onotone}) together with (\ref{usefulfact}).

Since  $\langle X_\epsilon(x)^2\rangle =\log(1/\epsilon)$, we have $\normord{e^{2 b X_\epsilon(x)}}=e^{2 b X_\epsilon(x)-2b^2 \log(1/\epsilon)}$.   Hence the cutoff version of $M_f(A)$ is
\be\label{mreg} M_{f,\epsilon}(A) =\int_A\d ^2x  f(x) e^{2b X_\epsilon(x) -2b^2\log (1/\epsilon)}.\ee
This is a well-defined random variable, and we want to investigate its convergence as $\epsilon\to 0$.

 \subsection{Martingales}\label{martingale}
 
 The circle cutoff that was just introduced will be useful in section \ref{detailedarguments}. 
 However, the  simplest proof of existence of limits such as $\lim_{\epsilon\to 0} M_{f,\epsilon}(A)$ actually requires a different sort of cutoff.
 
 In probability theory, a random variable $\Phi(t)$ that depends on a real 
 (or integer-valued) parameter $t$ is called a {\it martingale} if, conditional on a knowledge of  the whole history of  $\Phi(t)$ for $t\leq t_0$,
 the expected value of $\Phi(t)$ for any $t>t_0$ is precisely $\Phi(t_0)$.   It does not matter when in the past the process started;
 $\Phi(t)$ is a martingale if the expected value of $\Phi(t)$ for any $t>t_0$ is precisely $\Phi(t_0)$, independent of what happened for $t<t_0$.  For example, one's stake in a fair
 game of chance in which on the average one neither gains nor loses by playing is a martingale.   Regardless of one's stake at time $t_0$, the expected value of that stake at any
 future time is the same.
 
With a different regularization from the one that we described in section \ref{normalordered}, the cutoff dependence
of the random variable $M_f(A)$ is a martingale.   This was exploited in the foundational work of Mandelbrot \cite{Mandelbrot} and 
Kahane \cite{Kahane} on random variables such as $M_f(A)$.

Folllowing \cite{DS}, we will explain a variant of the construction that is slightly simpler to describe first for 
Liouville theory on a disc $D$, rather than on a two-sphere.
In this case, one has to study the same Gaussian free field $X(x)$ but now with, for example,  Dirichlet boundary conditions on the disc.    Let $f_i(x)$ be the
normalized eigenfunctions of the Dirichlet Laplacian on  $D$, with increasing eigenvalues $0<\lambda_1\leq \lambda_2\leq \cdots $.    The field $X(x)$ can be expanded in modes:\footnote{Kahane's procedure was slightly more complicated, for a reason that will be explained in section \ref{kahane}.}
\be\label{modexp}X(x)=\sum_{i=1}^\infty \frac{f_i(x) X_i}{\sqrt{\lambda_i}}, \ee
where $X_i$ are independent centered Gaussian variables, $\E[X_i X_j]=\delta_{ij}$.   The normal ordered exponential of $X(x)$
is formally an infinite product:
\be\label{nodexp} \normord{e^{2b X(x)}} =\prod_{i=1}^\infty e^{2b f_i(x)X_i -2b^2 f_i(x)^2}. \ee
For a regularization, we can simply truncate this as a finite product and define
\be\label{truncexp}\normord{e^{2b X(x)}}_n=\prod_{i=1}^n  e^{2b f_i(x)X_i -2b^2 f_i(x)^2}. \ee
The sequence of random variables $\normord{e^{2b X(x)}}_n$, $n=1,2,3,\ldots$, is a martingale, since conditioning on the values of $X_1,X_2,\ldots,X_n$ does not give any information
about $X_{n+1}$ and  $\E[e^{2b f_{n+1}(x)X_{n+1}-2b^2 f_{n+1}(x)^2}]=1$ for all $n$.  So, conditional on the values of $X_1,X_2,\ldots, X_n$, the expected value of $\normord{e^{2b X(x)}}_{n+1}$ is just the value of $\normord{e^{2b X(x)}}_{n}$.   Integration over a region $A\subset D$ does not disturb the martingale property, so likewise
the sequence of random variables
\be\label{zelob} M_{f,n}(A)=\int_A \d^2 x \,f(x) \normord{e^{2b X(x)}}_n \ee
is a martingale.  Of course, this is a positive martingale as $\normord{e^{2b X(x)}}_{n}$ is a product of real exponentials.
We have
\be\label{elob}\E[M_{f,n}(A)]=\int_A \d^2 x\, f(x)\ee
for all $n$, as this is a universal property of the normal ordered exponential of a Gaussian random variable.

We can make the same construction for Liouville theory on a two-sphere, with only slightly more care.   First we make  a Fourier expansion  $X(r,\theta)=\sum_{n\in \Z} X_n(r) e^{\i n\theta}$, as in eqn.~(\ref{fourier}).   The modes $X_n$ for
$n\not=0$ can be expanded in eigenfunctions of the Laplacian on the sphere, just as before.   For $n=0$, we just have to note that because of the constraint $X_0(r)=0$ at $r=1$,
the field $X_0(r)$ should be treated separately on the two discs $r\leq 1$ and $r\geq 1$. On each disc, $X_0(r)$ can be expanded in eigenmodes of the Dirichlet Laplacian on the disc.

A simple but rather striking result in probability theory is Doob's martingale convergence theorem (see \cite{Doob}, p. 319), which
says that if $\Phi(t)$ is a positive martingale  then with probability 1, $\Phi(t)$ converges to a nonnegative 
real number.  This limit is a random variable.    An informal
explanation of the proof is as follows.  Think of $\Phi(t)$ as the price of a commodity at time $t$.   The fact that $\Phi(t)$ is a martingale means that there is no way to
make an expected profit by buying and selling the commodity, because at any time $t_0$, the expected value of the price $\Phi(t_1)$ at any future time $t_1$ is just
the current price $\Phi(t_0)$.   The positivity of $\Phi(t)$ together with the martingale condition implies that $\Phi(t)$ does not have a positive
probability to diverge to $+\infty$ for $t\to\infty$; if this occurs, the expectation value of $\Phi(t)$ would also diverge as $t\to\infty$, contradicting the martingale property. So with
probability 1, $\Phi(t)$ does not diverge to $+\infty$.   
Suppose now that with probability $\delta>0$, $\Phi(t)$ does not converge.   This means that there
are numbers $a<b$ such that 
with positive probability, $\Phi(t)$ oscillates infinitely many times between $a$ and $b$.   In this case, a strategy of selling when the price is $b$ and buying when the price is $a$
has an average  gain of $+\infty$, contradicting the fact that there is no winning strategy.   So it must be that $\Phi(t)$ converges with probability 1 and this limit is automatically
a positive (or more precisely, non-negative) random variable.

Therefore, the random variables $M_{f,n}(A)$ have limits for $n\to\infty$.   These limits are nonnegative random variables and it makes sense to define their non-integer moments
(which in general are positive real numbers or $+\infty$).
However, there is some fine print in the martingale convergence theorem.   It is possible for the limit of a positive martingale to vanish.   This is possible even if the
martingale is normalized in the sense of eqn.~(\ref{elob}), which at first sight may appear to imply that the large $n$ limit of $M_{f,n}(A)$ cannot possibly vanish.

It is not difficult to give an artificial example of this behavior.\footnote{Let $\Phi_n$, $n=0,1,2,\ldots$ be a sequence of random variables, defined by saying that $\Phi_0=1$, and, inductively, if $\Phi_n=0$ then
 $\Phi_{n+1}=0$, but if $\Phi_n=2^n$,
 then $\Phi_{n+1} = 2^{n+1}$ with probability 1/2 and $\Phi_{n+1}= 0$ with probability $1/2$.   So  $\Phi_n$ equals $2^n$ with probability $2^{-n}$, and vanishes otherwise.
  We can think of $\Phi_n$ as the stake at time $n$ of
 a gambler who starts with \$1 at time 0, and in each time step, risks the entire stake on the flip of a coin, doubling the stake if the coin comes up heads and losing everything
 (and leaving the game) if it comes up tails.  Continuing to play the game does not change the expectation value of the gambler's stake, regardless of what that stake is at a given time, so this sequence of random variables is
 a martingale. The expectation value of the gambler's stake is \$1 at any time, but with probability 1, the gambler will ultimately lose the entire stake.
 So  even though $\E[\Phi_n]=1$ for all $n$, the large $n$ limit of $\Phi_n$ is~$0$.   
}   However, an example based on Brownian motion is more relevant to Liouville theory and also gives some intuition about how the infinite product (\ref{nodexp}) could converge to $0$.

First let us explain how Brownian motion is related to Liouville theory.
Set $\epsilon=e^{-t}$,  and consider the dependence on $t$ of  $X_\epsilon(x)=X_{e^{-t}}(x)$, for some fixed value of $x$.   We claim that the dependence
on $t$ of this variable is described by  Brownian motion -- the continuum version of a  random walk.   To see this, pick
 polar coordinates $r,\theta$ centered at $x_0$, so $x=x_0+r e^{\i\theta}$.   Let $r=e^{-t}$.   In terms of $t$ and $\theta$ the Gaussian free field 
 action for $X$ (see eqn.~(\ref{quadact})) 
 becomes\footnote{This is a consequence of the fact that the Gaussian free field action is invariant under a Weyl rescaling of the metric $g\to e^h g$ for any real-valued function $h$,
 together with the fact that the change of variables from Euclidean coordinates on the plane to polar coordinates $r,\theta$ is a conformal mapping.} 
 \be\label{newform2} I_\red=\frac{1}{4\pi}\int\d t\int_{0}^{2\pi} \d\theta \left(\left(\frac{\partial X}{\partial t}\right)^2 +\left(\frac{\partial X}{\partial\theta}\right)^2\right).\ee
 Expand $X(t,\theta)$ in Fourier modes:   \be\label{fourier} X(t,\theta)=\sum_{n\in\Z} X_n(t) e^{\i n\theta}. \ee
 The zero-mode $X_0(t)$ is precisely the variable $X_{e^{-t}}(x)$ whose dependence on $t$ we claim is described by Brownian motion.
The action $I_\red$ has the property that the  $X_n(t)$ for different values of $n$ are independent of
 each other.   The zero-mode  $X_0(t)$ is described by the action
 \be\label{zeroform} I_{\red,0}=\frac{1}{2} \int \d t \left(\frac{\d X_0(t) } {\d t}\right)^2. \ee
 This is precisely the action that describes Brownian motion, confirming the claim that Brownian motion governs   the dependence on $t$ of $X_0(t) = X_{e^{-t}}(x_0)$.
 
 As a matter of terminology, standard Brownian motion is Brownian motion that starts at the origin at time $0$, so $X_0(0)=0$, and satisfies $\E[X_0(t)^2]=t$.   Eqn.~(\ref{zeroform}) is normalized to lead to this time-dependence of the variance.  If we choose $x=0$, then  the constraint $X_0(0)=0$ is a restatement  of eqn.~(\ref{condx}), so in that case $B_t=X_0(t)$ is described by a standard Brownian motion.

Now let us go back to martingales.
The Brownian motion $X_0(t)$ is itself a fairly obvious example of a martingale.    $X_0(t)$ could have any value at time $t=t_0$, but the change in $X_0$ for time $t>t_0$ is
 independent of what happened for $t<t_0$, and has average value zero, so conditional on knowing $X_0(t_0)$, the average value of $X_0(t_1)$, for any $t_1>t_0$, is just $X_0(t_0)$.   
 
 Less obvious is that the normal-ordered exponential of $X_0(t)$  is again a martingale.    This will be used in section \ref{construction}.
 To show that $e^{\gamma X_0(t)-\frac{\gamma^2}{2}\E[X_0(t)^2]}$ is
 a martingale  for any $\gamma$, note that
 \be\label{compar4} e^{\gamma X_0(t_1)-\frac{\gamma^2}{2}\E[X_0(t_1)^2]}= U\cdot  e^{\gamma X_0(t_0)-\frac{\gamma^2}{2}\E[X_0(t_0)^2]},\ee
 with 
 \be\label{udef} U=    e^{\gamma (X_0(t_1)-X_0(t_0))-\frac{\gamma^2}{2}(\E[X_0(t_1)^2]-\E[X_0(t_0)^2])}.\ee
So  we have to show that conditional on a knowledge of $X_0(t_0)$ (which is equivalent to a knowledge of $e^{\gamma X_0(t_0)-\frac{\gamma^2}{2}\E[X_0(t_0)^2]}$), the expected
 value of $U$  is equal to 1.  
To show this,  we observe that $\gamma (X_0(t_1)- X_0(t_0))$ is a centered Gaussian random variable,
 so \be\label{comptn} \E[U]= e^{\frac{\gamma^2}{2}\E[(X_0(t_1) -X_0(t_0))^2]- \frac{\gamma^2}{2}(\E[X_0(t_1)^2]- \E[X_0(t_0)^2])}
=e^{-\gamma^2  \E[(X_0(t_1)-X_0(t_0)) X_0(t_0)]  }.\ee
 But $\E[(X_0(t_1)-X_0(t_0)) X_0(t_0)]=0$, since in Brownian motion, regardless of the value of $X_0(t_0)$, the expected value of the increment $X_0(t_1)-X_0(t_0)$ vanishes.

Since $e^{\gamma X_0(t)-\frac{1}{2}\gamma^2 \E[X_0(t)^2]}=e^{\gamma X_0(t)-\frac{1}{2}\gamma^2t} $ is  a positive martingale, it will have a large $t$ limit.    One might think that this limit would have to be nonzero,
since $\E[e^{\gamma X_0(t)-\frac{1}{2}\gamma^2 t}]=1$ for all $t$.   
  To see that this is wrong,
one can ask what is the typical value of the exponent at large $t$.   The first term $\gamma X_0(t)$ is typically of order $\gamma |t|^{1/2}$ while the second term $-\frac{\gamma^2}{2} t$ is negative
and proportional to $t$.   So a typical value of  $e^{\gamma X_0(t)-\frac{\gamma^2}{2} t}$ is exponentially small at large $t$.
  Although it is true that $\E[ e^{\gamma X_0(t)-\frac{\gamma^2}{2} t} ]=1$ for all $t$, this nonvanishing value at late times reflects
the contribution of exponentially unlikely events in which $X_0(t)$ is unexpectedly large.   For large $t$, with a probability that is exponentially close to 1, $e^{\gamma X_0(t)-\frac{\gamma^2}{2} t}$ is exponentially small, and the large $t$ limit of this variable vanishes.  The fact that the normal ordered exponential of Brownian motion decays exponentially in time will be important in section \ref{divergence}.

In more detail,
one way to prove that a positive random variable $\Phi$ vanishes is to find $p>0$ such that $\E[\Phi^p]=0$.   This implies that the probability density function $\rho$ of $\Phi$ satisfies
$\int_0^\infty \d x \rho(x) x^p=0$, forcing $\rho(x)=\delta(x)$ and $\Phi=0$.   Let us implement this with $\Phi(t)=e^{\gamma X_0(t)-\frac{\gamma^2}{2} t}$.
We have $\Phi(t)^p=e^{p\gamma X_0(t)-p\frac{\gamma^2}{2} t},$ leading to $\E[\Phi(t)^p]=e^{-\frac{\gamma^2 t}{2}(p-p^2)}$.   For $0<p<1$, the exponent is negative
and $\lim_{t\to\infty}\E[\Phi(t)^p]=0$, implying that $\lim_{t\to\infty}\Phi(t)=0$.     Such behavior actually occurs in Liouville theory for $b>1$,
as we will show in sections \ref{breakdown} and  \ref{triviality} by studying the $p^{th}$ moment of $M_f(A)$ for $p$ slightly less than 1.

Although $\normord{e^{2b X_\epsilon(x)}}$ for fixed $x$
can thus be viewed as a martingale in its dependence on $\epsilon$, if we integrate this quantity over a region $A\subset \Sigma$,
we do not get a martingale, because circles  centered at different points in $A$ intersect each other and the data on those circles are not independent.
That is why we explained the alternative regularization based on a normal mode expansion, which does give a martingale.   However, the regularization based on circle averaging has
other advantages and is used in the rest of this article.  Of course, physically one expects that the different regularizations will lead to equivalent results.

Thus a key issue in Liouville theory is to show that the random variables $M_f(A)$, defined by normal ordering, are actually nonzero for suitable $b$.
As we explain next, there is a relatively simple proof of this for $b<1/\sqrt 2$ with a suitable condition on $f$.  This leads to
 a simple rigorous definition of the Liouville correlation functions, for such  $b$, at least if the Liouville momenta $\alpha_i$ of the operators considered are small enough. 
 The proof that the $M_f(A)$ are nonzero for all $b<1$ is considerably more difficult and is deferred to section~\ref{construction}.

\subsection{$L^2$ Convergence for $b<1/\sqrt 2$}\label{ltwo}

Going back to the circle regularization, here  we will analyze the convergence of 
$M_{f,\epsilon} (A)$ as $\epsilon\to 0$ for the case $b<1/\sqrt 2$.  We  assume a region $A$
in which the two-point function is just $\E[X(x)X(y)]=\log(1/|x-y|)$.
To begin with, we  assume that $f$ is smooth and bounded; then we will consider singularities due to insertions of primary fields.   

Real-valued random variables that have convergent second moments form a vector space, because as $(\Phi+\Phi')^2\leq 2(\Phi^2+\Phi'^2)$, it follows
that if $\Phi$ and $\Phi'$ have convergent second moments, then so does $\Phi+\Phi'$:   $\E[(\Phi+\Phi')^2]\leq \E[2(\Phi^2+(\Phi')^2)]=2\E[\Phi^2]+2\E[(\Phi')^2]<\infty$.
This vector space has a positive-definite inner product $(\Phi',\Phi)=\E[\Phi'\Phi]$.   Completing it with respect to this inner product, we get a Hilbert space $\H$ of random
variables that have finite second moments.   We can think of this as the Hilbert space of square-integrable functions on some measure space.    If $\Phi_\epsilon$, for $\epsilon>0$,
is a family of vectors in $\H$, and 
the limit $\Phi=\lim_{\epsilon\to 0}\Phi_\epsilon$ exists in $\H$, then we say that the random variables $\Phi_\epsilon$
converge in $L^2$ to $\Phi$ as $\epsilon\to 0$.   Since a Hilbert space is complete by definition, 
the family $\Phi_\epsilon$ has such a limit as $\epsilon\to 0$ if and only if it is a Cauchy sequence in the $L^2$ sense, meaning that
\be\label{cauchy} \lim_{\epsilon,\epsilon'\to 0}\E[(\Phi_\epsilon-\Phi_{\epsilon'})^2]=0. \ee
We will show that for $b<1/\sqrt 2$, the family $M_{f,\epsilon}(A)$ is such a Cauchy sequence, and therefore the limit
\be\label{limobs}M_f(A)=\lim_{\epsilon\to 0} M_{f,\epsilon}(A) \ee
exists in the $L^2$ sense. 
Consequently, in this range of $b$, the second moment of $M_f(A)$, which is its norm squared as a Hilbert space vector, exists and is the
limit of the second moment of $M_{f,\epsilon}(A)$:
\be\label{secondmoment} \E[M_f(A)^2]= \lim_{\epsilon\to 0} \E[M_{f,\epsilon}(A)^2]. \ee
That in particular will show that $M_f(A)$ is not zero in this range of $b$.

It should not come as a surprise that $M_f(A)$ has a convergent second moment if $b<1/\sqrt 2$, because in the introduction we explained that an observable such as $M_f(A)$ (with smooth
bounded $f$) 
has a convergent  $n^{th}$ moment, for integer $n$,  if and only if $n<1/b^2$, and here we are simply making this statement for $n=2$.   What we gain by defining $M_f(A)$ as
a  random variable is that we can deduce that its non-integer moments exist, at least in a suitable range of exponents.   Indeed, if $\Phi$ is a positive random variable with 
$\E[\Phi^p]<\infty$ and
associated to a probability density $\rho$ on the positive half-line,
then for $0\leq q<p$, we have
\begin{align}\label{fracmom}\E[\Phi^q]&=\int_0^\infty \d u \rho (u)u^q=\int_0^1\d u \rho(u)u^q+\int_1^\infty \d u\rho(u) u^q\cr& \leq \int_0^1\d u\rho(u)+\int_1^\infty \d u \rho(u)u^p
\leq 1+\E[\Phi^p]<\infty. \end{align}
Setting $p=2$, convergence of $\E[\Phi^2]$ implies that the $q^{th}$ moment of $\Phi$  is always defined and convergent at least for $0\leq q\leq 2$.   In general, for $q<0$, the $q^{th}$ moment of a positive random variable $\Phi$ with convergent second moment will not necessarily converge;
it  will converge if and only if $\rho(u)$ behaves sufficiently well near $u=0$.   However,  in the particular case of  the random variables $M_f(A)$ in Liouville theory,  in order
for $M_f(A)$ to be extremely small requires
$X(x)$ to be very negative throughout the region $A$.   Because of the constraint that the average of $X$ on the unit circle vanishes, the Gaussian free field action $I_\red(X)$
becomes extremely large if $X(x)$ is very negative in a region $A$.   As a result, the measure $\rho(u)$ vanishes faster than any power of $u$ for $u\to 0$ and the moments
$E[M_f(A)^q]$ are convergent also for negative $q$.   Thus $M_f(A)$ has a $q^{th}$ moment for all $q\leq 2$, including $q<0$.  To learn what happens for 
$q>2$ requires more precise arguments that are presented in sections \ref{breakdown} and  \ref{detailedarguments}.

In view of eqn.~(\ref{secondineq}), we have 
\be\label{twocutoff} \E[\normord{e^{2bX_\epsilon(x)}} 
\, \normord{e^{2b X_\epsilon(y)}}] \leq{\rm {min}}(|x-y|^{-4b^2},\epsilon^{4b^2}),\ee
with equality if $|x-y|\geq 2\epsilon$.
This implies that
\be\label{convergent} \lim_{\epsilon\to 0} \E[M_{f,\epsilon}(A)^2]=\int \d^2x \d^2 y  f(x) f(y) \frac{1}{|x-y|^{4b^2}},\ee
since the region with $|x-y|<2\epsilon$ makes a negligible contribution in the limit $\epsilon\to 0$.   The right hand side is what one would naively write as the two-point function
of $M_f(A)=\int_A\d^2x  f(x) \normord{e^{2bX(x)}}$.   The more subtle fact that we are in the process of learning is that the quantity $M_f(A)$ that has this second moment can be defined
as a random variable that has non-integer moments.

To show that the limit $\lim_{\epsilon\to 0} M_{f,\epsilon}(A)$ exists in the $L^2$ sense, we need to show that the family of random variables $M_{f,\epsilon}(A)$ is Cauchy in the $L^2$
sense, meaning that 
\be\label{dlim} \lim_{\epsilon,\epsilon'\to 0} \E[(M_{f,\epsilon}(A)-M_{f,\epsilon'}(A))^2]=0.\ee   
Concretely the left hand side of eqn.~(\ref{dlim}) is 
\be\label{concreteform}   \int \d^2x \d^2y  f(x) f(y) \E[ (\normord{e^{2bX_\epsilon(x)}}-\normord{e^{2b X_{\epsilon'}(x)}})  (\normord{e^{2bX_\epsilon(y)}}-\normord{e^{2b X_{\epsilon'}(y)}}) ],\ee
and we need to show that this vanishes for $\epsilon,\epsilon'\to 0$, say with $\epsilon>\epsilon'$.  The integrand in (\ref{concreteform})
vanishes if $|x-y|>2\epsilon$ by virtue of eqn.~(\ref{uselater}),  while for $|x-y|<2\epsilon$, this integrand is a sum of four terms, each of which is bounded above by  $|x-y|^{-4b^2}$.
Since $\int_{|x-y|<2\epsilon} \d^2x |x-y|^{-4b^2}\sim \epsilon^{2-4b^2}$,  
 the integral vanishes for $\epsilon,\epsilon'\to 0$ if $b<1/\sqrt 2$, as we aimed to show.   The condition on $b$ that makes the contribution
from the region $|x-y|<2\epsilon$ vanish as $\epsilon\to 0$ is the same as the condition that ensures convergence of  the integral in (\ref{convergent}).

Up to this point, we have assumed that $f$ is smooth and bounded.   If instead we assume that $f$ has a singularity $f(x)\sim \frac{1}{|x-x_0|^{4b\alpha}}$ for some $x_0\in A$,
only one thing changes in the preceding analysis.   In order to have  $\lim_{\epsilon\to 0} \E[M_{f,\epsilon}(A)^2]<\infty$, we need 
\be\label{newint} \int_{A\times A}\d^2x \d^2y \frac{1}{|x-x_0|^{4b\alpha}}\frac{1}{|y-x_0|^{4b\alpha}}\frac{1}{|x-y|^{4b^2}}<\infty.\ee
Convergence of this integral for $x,y\to x_0$ is again equivalent to the condition that ensures that the family $M_{f,\epsilon}(A)$ is Cauchy for $\epsilon\to 0$.
This convergence
gives a new condition $8b\alpha+4b^2<4$ or
\be\label{alphabound}\alpha<\frac{1}{2b}-\frac{b}{2}.  \ee  This is a stronger condition than the Seiberg bound $\alpha\leq \frac{Q}{2}=\frac{1}{2b}+\frac{b}{2}$.
If the condition ({\ref{alphabound}) is satisfied, then the limit $M_f(A)=\lim_{\epsilon\to 0} M_{f,\epsilon}(A)$ exists in the $L^2$ sense
and in particular
\be\label{secondmmt} \E[M_f(A)^2]=\lim_{\epsilon\to 0}\E[M_{f,\epsilon}(A)^2].\ee 
We stress that the formulas for second moments obtained this way are not new.   What we gain from this discussion is only a framework in which noninteger moments of $M_f(A)$
make sense.   

Finally we can draw a conclusion about the existence of Liouville correlation functions.  
According to eqn.~(\ref{newform}), the Liouville correlator $\left\la \prod_{i=1}^n e^{2\alpha_i \phi(x_i)}\right\ra$ is the $w^{th}$ moment of $M_f=\int_\Sigma \d^2x  f(x) e^{2bX (x)}$,
with $w=Q-\sum_i\alpha_i$.   The arguments that we have given up to this point show that the Liouville correlation functions are well-defined if $b<1/\sqrt 2$,
 $\alpha_i<\frac{1}{2b}-\frac{b}{2}$ for all $i$, and $w<2$ or equivalently $Q-\sum_i\alpha_i<2$.   This is a nontrivial region of existence of the Liouville correlation 
 functions, but it is not the best possible.   To get the best possible results, one has to consider convergence in $L^p$ for $p<2$, rather than convergence in $L^2$.
 We explain the relevant arguments in section \ref{construction}.   
 
 First, though, we explain some relatively simple arguments showing that if certain inequalities are violated, these methods can {\it not} be used to define
 Liouville correlation functions.    In section
 \ref{detailedarguments}, it will become clear that these  inequalities are optimal.
 
 \subsection{Where Things Break Down}\label{breakdown} 
 
 In this section, we will explain that the probabilistic approach to Liouville theory that we have been describing fails if $b>1$ or if
 certain inequalities on the moments considered are violated.   The strategy we follow is  similar to arguments in the mathematical literature -- for example, see the proof of 
 Proposition 3.1 in \cite{RV} -- except that, deferring the rigorous arguments to section \ref{detailedarguments},  we explain how the key steps can be carried out using   an operator product expansion.
   
 We will need to make an operator product expansion in an unfamiliar situation with non-integer powers of the interaction, so perhaps we should recall the logic behind the operator
 product expansion.   The idea of the operator product expansion is that any observable that can be defined in a small region $A$ of spacetime can be expanded,
 in the limit that $A$ becomes small, as an asymptotic series in local operators that can be defined in region $A$.   The dominant contribution comes from the lowest dimension
 operator in region $A$ whose symmetry properties enable it to appear in the expansion.   In our application, the operators that can appear in the expansion are those with the right value
 of  what is sometimes called the Liouville momentum, which is defined so that the operator $\normord{e^{2\alpha X(x)}}$ or the product of this with any polynomial in derivatives of $X$, has Liouville momentum $\alpha$.      The reasoning behind the operator product expansion
 is robust enough that one expects it to be valid, in a reasonable quantum field theory, for any reasonable observable that can be defined in
 region $A$, including the somewhat exotic observables that we will encounter.
 
 As a first example of how the probabilistic approach to Liouville theory can break down, we will explain that the random distribution
  $\normord{e^{2bX(x)}}=\lim_{\epsilon\to 0}\normord{e^{2bX_\epsilon(x)}}$ vanishes if $b>1$.   Of course, regardless of the
 value of  $b$, the ``operator''
 $\normord{e^{2bX(x)}}=\lim_{\epsilon\to 0} \normord{e^{2bX_\epsilon(x)}}$ has meaningful correlation functions, for instance  $\E[\normord{e^{2bX(x)}}]=1$.   If $b>1$, however, the behavior of $\normord{e^{2bX_\epsilon(x)}}$ for $\epsilon\to 0$ is somewhat analogous
 to what happens
in the toy example described in section \ref{martingale} of the normal ordered exponential of a standard Brownian motion.   For small $\epsilon$, a random variable
$M_{f,\epsilon}(A)=\int_A\d^2x  f(x) \normord{e^{2bX_\epsilon (x)}}$ is close to zero with high probability and very large with a very small probability, in such a way that in the limit $\epsilon\to 0$, its
correlation functions remain nonvanishing  but the probability measure of  $M_{f,\epsilon}(A)$  converges to
a delta function at the origin.   
In such a situation, 
moments of $M_f(A)$ of non-integer order cannot be defined.

 As discussed in section \ref{martingale}, to show that a positive random variable $\Phi$ vanishes, it suffices to find a positive number $p$ such that $\E[\Phi^p]=0$.    In the case of $M_f(A)=\int_A \d^2x f(x)\normord{e^{2b X(x)}}$, we will show
 that the $p^{th}$ moment vanishes for $p$ slightly less than 1.  The precise range of $p$ for which the argument works will become clear.  Of course, once one has $\E[M_f(A)^p]=0]$
 (in the limit $\epsilon\to 0$) for some $p>0$, it follows that $M_f(A)=0$ as a random variable and all its moments vanish.
 
 First of all, for $\Phi_1,\Phi_2,\ldots, \Phi_n>0$ and $0<p<1$, one has \be\label{firstone} \left(\sum_i \Phi_i\right)^p\leq \sum_i \Phi_i^p.\ee
    Applying this to 
 $M_{f_i}(A)$ for a family of nonnegative functions $f_i$, and setting $f=\sum_i f_i$, we see that the
 $p^{th}$ moment of $M_{f}(A)$ will vanish if that is true for each $M_{f_i}(A)$.   So, writing any $f$ as a sum of functions supported in small squares,
  we can reduce to the case that the support of $f$ is a small square in $\Sigma$.
 We can also assume that $f$ is a constant function, since replacing $f$ by a constant $c>f$ will only make $M_f(A)$ bigger.   
 Since $M_c(A)$ is just proportional to $c$, we also lose nothing essential if we set $c=1$.

 So finally the only case that we have to consider is the random variable $M(A)=\int_A\d^2x e^{2b X(x)}$, where $A$ is a square in the complex $x$-plane.  Pick an
 integer $q$ and decompose $A$ as   the union of $q^2$ identical squares $A_1,\cdots , A_{q^2}$ that are each smaller by a factor of $q$.  Using eqn. (\ref{firstone}) again, we get
 \be\label{qineq} \E[M(A)^p]\leq \sum_{i=1}^{q^2} \E[M(A_i)^p]=q^2 \E[M(A_{(q)})^p], \ee
 where $A_{(q)}$ is any one of the little squares.    In the last step we use the fact that $\E[M(A_i)^p]$ is independent of $i$.
 
 We can choose $A_{(q)}$ so that as $q$ increases, $A_{(q)}$ converges to a point $x_0\in A$.   This is the setting of the operator product expansion: asymptotically for $q\to\infty$,
 we can expect to expand any observable or product of observables supported in $A_{(q)}$, such as $M(A_{(q)})^p=\left(\int_{A_{(q)}} \d^2x e^{2b X(x)}\right)^p $, in a series of local operators at the point $x_0$.
 The operator of lowest dimension with the correct value of the Liouville momentum is simply $\normord{e^{2bp X(x_0)}}$, so the expansion takes the form
 \be\label{expform}M(A_{(q)})^p\sim C(q) \normord{e^{2b pX(x_0)}}+\cdots,\ee
 where the corrections are subleading for large $q$.   A simple scaling argument determines the $q$ dependence of the coefficient $C(q)$.
 The operator $\normord{e^{2\alpha X(x)}}$ in the Gaussian free field theory  has holomorphic or antiholomorphic dimension\footnote{In the introduction, in a related calculation,
 we scaled using the Liouville theory formula for the dimensions of operators, namely $\h\Delta_\alpha=-\alpha^2+Q\alpha$.   Here, as we have eliminated the Liouville zero-mode
 and $X$ is just a Gaussian free field, it is more natural to scale using the Gaussian free field formula $\Delta_\alpha=-\alpha^2$.   The two formulas differ by whether
 a scaling of spatial coordinates is accompanied by a shift of the field.   Regardless, in either of the two calculations, the term linear in $\alpha$ cancels out and it does
 not matter which of the two formulas one uses.} 
 $\Delta_\alpha=-\alpha^2$ in energy units, so in two dimensions its
 overall scaling dimension is $-2\alpha^2$ in energy units  or $+2\alpha^2$ in length units.
 So $M(A_{(q)})=\int_{A_{(q)} }\d^2x\normord{ e^{2b X(x)}}$ has length dimension $2+2b^2$ (counting 2 for the integral) and $M(A_{(q)})^p$ has length dimension
 $(2+2b^2)p$.   On the right hand side of eqn.~(\ref{expform}), the operator $\normord{e^{2bp X(p)}}$ has length dimension $b^2p^2$.  The difference in dimensions between
 the left and right determines the $q$-dependence of $C(q)$.  Indeed, $1/q$ is a length parameter, since it determines the size of the region $A_{(q)}$.   So the power of $1/q$ is
 the difference of length dimensions between the two sides, and a more precise form
 of the expansion 
 (\ref{expform}) is
 \be\label{nexpform}M(A_{(q)})^p= c q^{2b^2p^2-(2+2b^2)p} \normord{e^{2b p X(x_0)}}+\cdots \ee
 where now the constant $c$ is independent of $q$ for $q\to\infty$ and again $+\cdots$ denotes subleading terms. Now we take the expectation value of eqn. (\ref{nexpform}),  using $\E[\normord{e^{2bp X(x_0)}}]=1$.  Replacing $c$ with a slightly larger constant $c'$ and taking $q$ large enough that the subleading terms are negligible, we learn from eqn. (\ref{expform}) 
 that
 \be\label{answer} \E[M(A)^p)]\leq  c' q^{2b^2 p^2-(2+2b^2)p +2}=c' q^{2(p-1)(b^2p-1)}.\ee
 For $b>1$ and $p$ slightly less than 1, $2(p-1)(b^2p-1)<0$, and therefore the validity of the inequality (\ref{answer}) for arbitrarily large $q$ implies
 that
 \be\label{vanishing} \E[M(A)^p]=0.\ee
   As explained previously, this implies that $M(A)=0$ for $b>1$ and therefore that the methods that we have been reviewing in this article
 do not suffice to define Liouville theory for such values of $b$.
 
 There is actually also a mathematical definition of Liouville theory at $b=1$.
 At that value of $b$, rather than a normal ordered version of $e^{2b X(x)}$, one considers \cite{DRSV}, as one might
possibly expect from the reasoning in \cite{Seiberg}, a normal ordered version of $-X(x) e^{2 X(x)}=-\frac{1}{2}\left. \frac{\partial}{\partial b}\right|_{b=1} e^{2b X(x)}$.   To be more precise, one considers \be\label{delm}-\frac{1}{2}\left.\frac{\partial}{\partial b}\right|_{b=1} \normord{e^{2b X(x)}}=-\frac{1}{2}\left.\frac{\partial}{\partial b}\right|_{b=1}
\lim_{\epsilon\to 0}e^{2b X_\epsilon(x)-2b^2\E[X_\epsilon^2]}=\lim_{\epsilon\to 0}\left[ \left(2\E[X_\epsilon(x)^2]-X_\epsilon(x)\right)e^{2X_\epsilon(x)-2 \E[X_\epsilon(x)^2]}\right].\ee
The DOZZ formula shows that the theory is actually holomorphic in $b$ (in the half-plane ${\rm Re}\,b>0$), so the region $b>1$ is related by analytic continuation
to the region $b<1$, and in fact there is a symmetry $b\leftrightarrow \frac{1}{b}$.
 But it is extremely unclear how  to define the theory directly for $b> 1$.    
Note that although $-\frac{1}{2} \frac{\partial}{\partial b} e^{2b X(x)}$ is negative classically, it turns
out to be positive quantum mechanically at $b=1$.

Henceforth we assume  $b<1$.   In section \ref{ltwo}, we were able to define moments $\E[M_f(A)^p]$ for a certain range of $p$ assuming $b<1/\sqrt 2$,
 and those results and limits will be improved in section \ref{detailedarguments}.   Here we will prove that the moments of $M_f(A)$ do {\it not} exist if $p$ is too large.
 The results  are best possible, as will become clear in section \ref{detailedarguments}, and are related to a nonperturbative pole of the DOZZ formula.
 
 First we will prove that $\E[M_f(A)^p]$ diverges if $p>1/b^2$.   This was already explained in the introduction for the case that $p$ is an integer.   The statement actually holds whether
 $p$ is an integer or not.
  
 Since $M_f(A)$ is nonnegative for any nonnegative $f$ and can only increase as $f$ increases, to show that it diverges, it suffices to show that it diverges for the case that $A$ is a small
 square and $f$ is a constant function, which we may as well take to be $f=1$.   Again we decompose $A$ as the union of $q^2$ little
 squares $A_i$ that are each translates of $[0,1/q]^2$.  Since we assume $b<1$ and we want to investigate what happens for $p>1/b^2$, we can assume $p>1$.
 Since 
 \be\label{simplineq} \biggl(\sum_i\Phi_i\biggr)^p\geq \sum_i \Phi_i^p\ee for $\Phi_i\geq 0$ and $p>1$, we have an inequality similar to that of eqn.~(\ref{qineq}) but going in the opposite
 direction:
 \be\label{qineq2} \E[M(A)^p]\geq \sum_{i=1}^{q^2} \E\left[\sum_i M(A_i)^p\right]=q^2 \E\left[M(A_{(q)})^p\right], \ee
 where again $A_{(q)}$ is one of the little squares.   For large $q$, we again make the operator product expansion (\ref{nexpform}), leading to the same inequality as in
 (\ref{answer}) but in the opposite direction:
 \be\label{answer2} \E[M(A)^p)]\geq c q^{2b^2 p^2-(2+2b^2)p +2}.\ee
 From the fact that this inequality holds for arbitrarily large $q$, we learn that $\E[M(A)^p]=+\infty$  if the exponent $2(p-1)(b^2p-1)$ is positive.   Given that $p>1$ (as assumed in the
 derivation), this inequality is satisfied precisely if $p>1/b^2$, so the moments diverge in that range of $p$.   
 
Now, let us assume that $p>1$ is
 such that the moments converge when $f$ is bounded, and ask what happens in the presence of a singularity\footnote{We will consider the case of just one such singularity;
 if there are multiple singularities, they can be treated independently by the same reasoning that we are about to explain.}  in $f$ created by one of the Liouville vertex operators, say
 $f=|x|^{-4b\alpha}$ for some $\alpha$ near $x=0$.  For any $\eta>0$, we can write
 \be\label{decomp} M_f(A)=\int_{|x|\leq \eta}\d^2x f(x) \normord{e^{2b X(x)}}+\int_{|x|\geq \eta} \d^2x f(x) \normord{e^{2bX(x)}}. \ee
 Because for $\Phi_1,\Phi_2>0$, $(\Phi_1+\Phi_2)^p\leq 2^p (\Phi_1^p+\Phi_2^p)$, to get a divergence in $\E[M_f(A)^p]$, there must be a divergence
 if $M_f(A)$ is replaced by one of the two terms on the right hand side of eqn.~(\ref{decomp}).    Since we assume that $p$ is such that the $p^{th}$ moment of $M_f(A)$
 converges if $f$ is bounded, 
the second term on the right hand side of eqn.~(\ref{decomp}) has  convergent $p^{th}$ moment and any divergence will come from the first term. 
 Similarly, adding to $f$ a bounded
 function will not affect whether there is a new divergence in the $p^{th}$ moment associated to the singularity in $f$.   Hence, we can assume that $f$ is precisely equal to
 $|x|^{-4b\alpha}$ in some disc around the origin, say of radius $\eta_0$.
Since restricting to a smaller region will make $M_f(A)$ smaller, we have  for any $\eta\leq \eta_0$
\be\label{releq} \E[M_f(A)^p]\geq \E[M_\eta^p],\ee
where 
\be\label{meta} M_\eta=\int_{|x|\leq \eta}\d^2x |x|^{-4b\alpha} \normord{e^{2bX(x)}}.\ee
So we will have $\E[M_f(A)^p]=+\infty$ if $\E[M_\eta^p]$ diverges for $\eta\to 0$ (in which case the same reasoning shows that  $\E[M_\eta^p]$ is infinite even before taking $\eta\to 0$).
 
We can use the operator product expansion to determine the behavior of $M_\eta^p$ for $\eta\to 0$.   
As $\eta\to 0$, $M_\eta^p$ is an observable supported in a very small neighborhood of $x=0$, so it has an asymptotic expansion in local operators
at $x=0$.   The lowest dimension such operator that has the same Liouville momentum as $M_\eta^p$ is $e^{2bp X(0)}$, with length dimension $2b^2p^2$.
Since the length dimension of $M_\eta^p$ is $p(2-4b\alpha +2b^2)$, where we include  2 from the integral and $-4b\alpha$ from the explicit factor $|x|^{-4b\alpha}$,
the form of the expansion is
\be\label{opexp} M_\eta^p\sim C \eta^{p(2-4b\alpha+2b^2)-2p^2b^2} \normord{e^{2bp X(0)}}+\cdots ,\ee
where $C$ is a constant, independent of $\eta$, and the omitted terms are subleading as $\eta\to 0$.     Taking expectation values, using $\E[\normord{e^{2bpX(0)}}]=1$, and replacing   $C$ by a slightly larger constant so that subleading terms in eqn.~(\ref{opexp}) can be dropped for sufficiently small  $\eta$, we get a more 
explicit version of the inequality (\ref{releq}):
\be\label{eleq} \E[M_f(A)^p]\geq C \eta^{p(2-4b\alpha+2b^2)-2p^2b^2}=C\eta^{2p(1-2b\alpha+b^2-pb^2)}.\ee
So $\E[M_f(A)^p]=+\infty$ when the exponent on the right hand side of this inequality is negative,
or in other words when
\be\label{newlimit}p>\frac{1}{b^2}+1-\frac{2\alpha}{b}. \ee
For integer $p$, this was deduced in the introduction.

As asserted in eqn. (\ref{mmtdegree}), Liouville correlation functions are moments of the Liouville interaction of order $w=(Q-\sum_i\alpha_i)/b$ (times some known factors), where 
$Q=b+1/b$. 
So the preceding results imply  that the Liouville correlators diverge, even after removing an explicit factor that contains the perturbative poles,  unless
\be\label{firstcond} \sum_i\alpha_i >b\ee
and
\be\label{secondcond} \sum_i\alpha_i >2\alpha_k~~{\rm {for~all}}~ k. \ee
These constraints are the strongest possible, since in
 section \ref{detailedarguments}, we will see that $\E[M_f(A)^p]<\infty$  if they are satisfied along with
 the Seiberg bound
$\alpha_i  < Q/2$.    

 Note that in the case of the three-point function, eqn.~(\ref{secondcond}) is a triangle inequality,
asserting that $\alpha_1+\alpha_2>\alpha_3$, and cyclic permutations.  In the semiclassical regime of small $b$, the triangle inequality has a semiclassical
interpretation, explained in \cite{HMW}, in terms of complex saddle points of the classical Liouville equation.

\subsection{Comparison to the DOZZ Formula}\label{dozzcompar}

The results (\ref{firstcond}) and (\ref{secondcond}) can be compared to the DOZZ formula, whose general form is as follows.
By virtue of the $SL(2,\C)$ symmetry of a Riemann surface of genus 0, the Liouville three-point function in genus 0 has the general form 
\be
\label{3point}
\left\langle\prod_{i=1}^3\normord{ e^{2\alpha_i(x_i)}}\right\rangle=
\frac{C(\alpha_1,\alpha_2,\alpha_3)}{|x_{12}|^{2(\h\Delta_1+\h\Delta_2-\h\Delta_3)}
|x_{13}|^{2(\h\Delta_1+\h\Delta_3-\Delta_2)}|x_{23}|^{2(\h\Delta_2+\h\Delta_3-\h\Delta_1)}},
\ee
where $\h\Delta_i=\alpha_i(Q-\alpha_i)$, $x_{ij}=x_i-x_j$, and $C(\alpha_1,\alpha_2,\alpha_3)$ does not depend on the $x_i$.
The DOZZ formula is a formula for the function $C(\alpha_1,\alpha_2,\alpha_3)$:
\begin{align}
\label{dozz} \nonumber
&C(\alpha_1,\alpha_2,\alpha_3)=\left[\pi \mu \gamma(b^2) 
b^{2-2b^2}\right]^{(Q-\sum{\alpha_i})/b}\\
&\times\frac{\Upsilon_0 \Upsilon_b(2\alpha_1)\Upsilon_b(2\alpha_2)
\Upsilon_b(2\alpha_3)}{\Upsilon_b(\alpha_1+\alpha_2+\alpha_3-Q)
\Upsilon_b(\alpha_1+\alpha_2-\alpha_3)
\Upsilon_b(\alpha_2+\alpha_3-\alpha_1)\Upsilon_b(\alpha_1+\alpha_3-\alpha_2)}.
\end{align}
Here $\Upsilon_b(x)$, for ${\mathrm Re}\,b>0$,  is an entire function of $x$ that has simple zeroes at (and only at) 
$x=-m/b-nb$ and $x=(m+1)/b+(n+1)/b$, for any non-negative integers $m,n$;
$\Upsilon_0=\left.\d\Upsilon_b(x)/\d x\right|_{x=0}$; and $\gamma(x)=\Gamma(x)/\Gamma(1-x)$.

Zeroes of the denominator of the DOZZ formula lead to 
poles of the Liouville three-point function.   In particular, the perturbative poles found by Goulian and Li \cite{GL} and described in the introduction 
 result from the zeroes of $\Upsilon_b(\alpha_1+\alpha_2+\alpha_3-Q)$
at $\sum_i\alpha_i-Q=-nb$.     The conditions that we have found in eqns. (\ref{firstcond}) and (\ref{secondcond}) correspond to some of the nonperturbative poles
of the DOZZ formula.   The condition (\ref{firstcond}) is associated to the pole of the DOZZ formula at $\sum_i\alpha_i -Q=-\frac{1}{b} $, and the condition (\ref{secondcond})
is associated to the poles of the DOZZ formula at $\alpha_1+\alpha_2-\alpha_3=0$ or permutations thereof.

After integrating over the Liouville zero-mode, the remaining integral over the Gaussian free field $X$ has an extended region of convergence, as remarked in the introduction.
This extended region is bounded by the conditions (\ref{firstcond}) and (\ref{secondcond}) along with the Seiberg bound $\alpha_i\leq Q/2$ for all $i$.  What happens at the Seiberg bound is completely different from what happens at the other boundaries.  At the Seiberg bound, the DOZZ formula has not a pole but a zero,
resulting from the zero of $\Upsilon_b(2\alpha)$ at $\alpha=0$.     We have interpreted the conditions (\ref{firstcond}) and (\ref{secondcond}) in terms of ultraviolet
divergences, which lead to poles of the DOZZ formula.   What can  account for a zero of the DOZZ formula?

In fact, suppose that we increase one of the Liouville momenta in the three-point function, say $\alpha_1$, until the Seiberg bound is violated.  To avoid the pole at $\alpha_1=\alpha_2+\alpha_3$,  we maintain the triangle inequality $\alpha_1<\alpha_2+\alpha_3$.   So by the time one reaches the Seiberg bound at $\alpha_1=Q/2$, we will have
$\sum_i\alpha_i>Q$ and therefore $w=(Q-\sum_i\alpha_i)/b<0$.  This leads to vanishing of the DOZZ formula for the following reason.
 The Liouville three-point function is, as in eqn. (\ref{newform}),
a multiple of $\E\left[\left(\int \d^2x f(x) e^{2bX(x)}\right)^w\right]$,  with $f(x)\sim |x-x_1|^{-4b\alpha_1}$ for $x\to x_1$.   
In the probabilistic approach that is reviewed in the present article, the interpretation of the Seiberg bound is that if $\alpha\geq Q/2$, then, for any $\eta
>0$, the quantity  $\int_{|x-x_1|<\eta}\d^2x |x-x_1|^{-4b\alpha}e^{2b X(x)}$  is equal to $+\infty$
with probability $1$.  For the proof of this assertion, see section \ref{divergence}.   Because this quantity  equals $+\infty$ with probability $1$, its negative
powers vanish with probability $1$, implying that  $\E\left[\left(\int \d^2x f(x) e^{2bX(x)}\right)^w\right]=0$ for $w<0$.   This is a probabilistic interpretation of the vanishing of the DOZZ
formula at the Seiberg bound.

The DOZZ formula shows that the Liouville correlation function  can be analytically continued past the Seiberg bound (as was actually anticipated in \cite{Seiberg}).
However, because the DOZZ formula has a simple zero at the Seiberg bound, the three-point function, continued so that one of the Liouville momenta slightly
exceeds $Q/2$, becomes negative, contradicting the positivity that appears to follow from the definition of the correlation function as $\la\prod_i\normord{ e^{2\alpha_i\phi(x_i)}}\ra$
and showing that the correlation functions analytically continued past the Seiberg bound do not have a simple probabilistic interpretation.  Positivity is similarly
lost upon analytic continuation past the simple poles that mark the other boundaries of the extended region of convergence of the path integral.   Beyond the other
boundaries of the extended region of convergence are the many other nonperturbative poles of the DOZZ formula, whose interpretation from a probabilistic point of view
is not clear.

\section{Liouville Theory Correlators: Detailed Arguments}\label{detailedarguments}

In the rest of this article, we will sketch  more detailed rigorous mathematical arguments about Liouville theory.
The arguments are mostly adapted from the existing literature, though the treatment of certain singular integrals in section \ref{singular} may be new.

First we explain two useful tools: the near scale invariance of the Gaussian free field (section \ref{nearscale})
and an inequality due to Kahane \cite{Kahane} (section \ref{kahane}).    Then in section \ref{triviality}, we show that in the limit $\epsilon\to 0$
the random variables $M_{f,\epsilon}(A)$ of Liouville theory vanish if $b>1$.  Such a result was originally obtained in \cite{Kahane} (with a different cutoff).
 Subsequently, we assume that $b<1$ and we consider the behavior of the moments $\E[M_{f,\epsilon}(A)^p]$ with $f$ assumed initially to be bounded.  In section \ref{moments}, we show
 that these moments diverge in the limit $\epsilon\to 0$ if $p>1/b^2$ but converge to finite, positive values if $1\leq p<1/b^2$ (also a result of \cite{Kahane}).   In section \ref{singular}, we consider
 the case that $f(x)$ has a power law singularity\footnote{\label{unfortunate}Unfortunately, there is a clash of notation between what has become standard in the physics
 and mathematics literatures on Liouville theory.   What has been called $\alpha$ in the math literature is $2\alpha$ in the physics literature.   To facilitate comparison of the following
 formulas with the mathematics literature, we define $\upalpha=2\alpha$, where the usual Liouville primary fields are denoted as $e^{2\alpha\phi(x)}$ in the physics
 literature and as $e^{\upalpha \phi(x)}$ in the math literature.   Up to this point, all formulas have been written in terms of $\alpha$; henceforth we use $\upalpha$.   Another detail is
 that, as  remarked
 in footnote \ref{conventions}, the mathematics literature is mostly written in terms of the variable $\gamma=2b$.}
  $1/|x-x_0|^{2\upalpha b}$ for some $\upalpha$ and again determine the range of $p$ for which $\E[M_{f,\epsilon}(A)^p]$ converges
 for $\epsilon\to 0$, as first analyzed in   \cite{KRV}.   In section \ref{divergence}, we show that $\lim_{\epsilon\to 0} M_f(A)=\infty$
 with probability 1 if $f$ has a singularity with $\upalpha$ exceeding the Seiberg bound at $\upalpha=Q$.   As already explained in section \ref{dozzcompar}, this accounts for the vanishing
 of the DOZZ formula at the Seiberg bound.  
 
 The results summarized up to this point do not address the question of whether, say for bounded $f$, the random variables $\E[M_{f,\epsilon}(A)]$ have nonzero limits for $\epsilon\to 0$.
 In general, a family of random variables $M_{f,\epsilon}(A)$ can converge to 0 for $\epsilon\to 0$, even while some moments remain nonzero.   As explained in section \ref{martingale},
 this can happen if, for very small $\epsilon$, the nonzero moments of $M_{f,\epsilon}(A)$ reflect the contributions of very rare events in which $M_{f,\epsilon}(A)$ is unexpectedly large.
 In section \ref{construction}, we show that this does not happen and that for all $b<1$, a nontrivial Liouville probability measure does exist in the limit $\epsilon\to 0$.   This is accomplished by
 showing that (for bounded $f$), the random variables $M_{f,\epsilon}(A)$ converge in $L^1$ to nonzero limits 
 as $\epsilon\to 0$.    In general, a positive martingale that converges in $L^1$ and has
 a convergent $p^{th}$ moment also converges in $L^p$, so it follows\footnote{Here we are tacitly assuming that $\lim_{\epsilon\to 0}M_{f,\epsilon}(A)$ is equivalent to
 a similar limit defined using Kahane's cutoff, which gives the martingale property.  This can plausibly be proved using Kahane's inequality, though we will not do so.} 
 that (for bounded $f$), $M_{f,\epsilon}(A)$ converges in $L^p$.

\subsection{Near Scale Invariance}\label{nearscale}

In section \ref{correlators}, we defined the Gaussian free field so that for $|y|,|y'|<1$, the two-point correlation function is
just
\be\label{logfn} \E[X(y) X(y')]=\log\frac{1}{|y-y'|}. \ee
This immediately implies that  for $q>1$,
 \be\label{compar}\E[X(y/q)X(y'/q)]=\log\frac{q}{|y-y'|} = \E[X(y) X(y')]+\log q. \ee
Thus,   $X(y/q)$ has the same distribution as $X(y)+\Omega_q$, where $\Omega_q$ is a Gaussian random variable of mean zero and variance $\log q$, and independent
 of $X(y)$.   Hence for any function $F$, 
 \be\label{turno} \E[F(X(y/q))]=\E[F(X(y)+\Omega_q)].\ee
 In general, we have to consider regularized versions of various expressions.   If instead of $X(y)$ we consider its circle regularized version $X_\epsilon(y)$, then a formula
 similar to eqn. (\ref{logfn}) holds, except that we have to  rescale $\epsilon$ along with the other lengths:
 \be\label{compar2}\E[X_{\epsilon/q}(y/q)X_{\epsilon/q} (y'/q)]= \E[X_\epsilon(y) X_\epsilon(y')]+\log q. \ee   Therefore the cutoff version of eqn. (\ref{turno})   is
 \be\label{murno} \E[F(X_{\epsilon/q}(y/q))]=\E[F(X_\epsilon(y)+\Omega_q)]. \ee
 In practice, since two different values of the cutoff parameter appear in  eqn.~(\ref{murno}),   to make use of this formula,  we frequently 
 need to know how to compare the theories with different
 cutoffs.   It turns out that this can be done quite effectively using eqn.~(\ref{monotonic}), which asserts that $\E[X_\epsilon(x) X_\epsilon(y)]$ is monotonically increasing as
 $\epsilon$ becomes smaller.    
 
 Kahane \cite{Kahane} considered  a Gaussian free field that has  a two-point function assumed to be of the form
 \be\label{kform}\E[X(y)X(y')]=\sum_{i=1}^\infty g_i(y,y'), \ee
where the functions $g_i(y,y')$ are positive pointwise and also are positive as kernels, in the sense that for real-valued functions $f$, the quadratic form $Q_i(f)=\int \d^2 y \d^2 y' f(y)g_i(y,y') f(y')$ is
positive-definite.   For suitable $g_i$, the logarithmic correlations of Liouville theory can be expressed as in eqn. (\ref{kform}).   We made a similar expansion in section \ref{martingale}, but in terms of the eigenfunctions $f_i(y)$ of the Laplace operator,
 which are not positive;  
 in Kahane's approach, positivity is important.  Positivity of the quadratic forms makes it possible to define Gaussian variables $X_{(i)}$ with $\E[X_{(i)}(y) X_{(i)}(y')]= g_i(y,y')$ 
 (and with $X_{(i)}$ independent of $X_{(i')}$ for $i\not=i'$), and then if one sets $X_n=\sum_{i=1}^n X_{(i)}$, one has 
  \be\label{kform2}\E[X_n(y)X_n(y')]=\sum_{i=1}^n g_i(y,y').\ee  Pointwise positivity  of the functions $g_i(y,y')$ ensures that  $\E[X_n(y)X_n(y')]$ is an increasing function of $n$, 
 analogous to the fact that with the cutoff based on circle averaging, 
   $\E[X_\epsilon(x) X_\epsilon(y)]$  monotonically increases   for $\epsilon\to 0$.
  
  So in either approach, the two-point function increases as the cutoff is removed.   This has important implications when combined with Kahane's inequality, which we describe
  next.
    
\subsection{Kahane's Inequality}\label{kahane}

Let $X_i$ and $Y_i$, for $i=1,\ldots, n$ be two families of independent, centered Gaussian random variables and let $c_i$, $i=1,\ldots, n$, be  positive numbers.\footnote{In explaining
 Kahane's inequality, we follow the appendix of \cite{RV}.}
Set  $Z_i(t)=\sqrt t X_i +\sqrt{1-t} Y_i$, $0\leq t\leq 1$.  Of course, for each $t$ the $Z_i$ are again independent Gaussian random variables.
 Let $\phi:\R_+\to \R$ be a twice differentiable function with at most polynomial growth at infinity, and define
\be\label{zoro}\varphi(t)=\E\left[\phi\left(\sum_i c_i \normord{e^{Z_i(t)}} \right)\right].\ee
Then a straightforward calculation reveals a surprisingly simple and useful result:
\be\label{boro}\frac{\d\varphi(t)}{\d t}= \frac{1}{2}\sum_{i,j=1}^n c_i c_j (\E[X_i X_j]-\E[Y_i Y_j])\E\left[\normord{e^{Z_i(t)+Z_j(t)}}\phi''\left(\sum_{k=1}^n c_k \normord{e^{Z_k(t)}}\right)\right].\ee
Suppose then that 
\be\label{reli} \E[X_i X_j]\geq  \E[Y_i Y_j],~~~1\leq i,j\leq n.\ee   Then if the function $\phi$ is convex, $\phi''\geq 0$, we have $\varphi(1)\geq\varphi(0)$ or
\be\label{beli} \E\left[\phi\left(\sum_i c_i \normord{e^{X_i}}\right)\right]\geq \E\left[\phi\left(\sum_i c_i \normord{e^{Y_i}} \right)  \right],\ee
while if $\phi$ is concave, $\phi''\leq 0$, we have the opposite inequality   
\be\label{zeli}  \E\left[\phi\left(\sum_i c_i \normord{e^{X_i}}\right)\right]\leq \E\left[\phi\left(\sum_i c_i \normord{e^{Y_i}}\right)\right].\ee
Specializing to moments,     
\be\label{weli}\E\left[\left(\sum_i c_i \normord{e^{X_i}} \right)^p\right]\geq          \E\left[\left(\sum_i c_i\normord{e^{Y_i}}\right)^p\right],~~ p\geq 1 \ee
and
\be\label{teli} \E\left[\left(\sum_i c_i \normord{e^{X_i}} \right)^p\right]
           \leq  \E\left[\left(\sum_i c_i \normord{e^{Y_i}}\right)^p\right],~~ 0\leq p\leq 1. \ee           
Equation (\ref{monotonic}) implies that $\E[X_{\epsilon'}(y) X_{\epsilon'}(y')]\geq \E[X_{\epsilon}(x) X_{\epsilon}(y)] $ if $\epsilon'\leq \epsilon$.   Hence a continuum limit of
the inequalities (\ref{weli}) and (\ref{teli}) gives for $\epsilon'\leq \epsilon$, and any non-negative function $f$,
\be\label{momentineq} \E\left[\left(\int_A\d ^2x f(x) \normord{e^{2b X_{\epsilon'}(x)}}\right)^p\right]   \geq    \E\left[\left(\int_A\d^2x f(x) \normord{e^{2b X_{\epsilon}(x)}}\right)^p\right],~~~p\geq 1\ee
and
\be\label{momentineq2} \E\left[\left(\int_A\d^2x f(x) \normord{e^{2b X_{\epsilon'}(x)}}\right)^p\right]   \leq    \E\left[\left(\int_A\d^2x f(x) \normord{e^{2b X_{\epsilon}(x)}}\right)^p\right],~~~0\leq p\leq 1.\ee        
Similar inequalities hold with Kahane's cutoff, since the cutoff two-point function (\ref{kform2}) is an increasing function of $n$.  Kahane's used these inequalities as follows.
There are many possible choices of the positive functions $g_i(x,x')$ in the decomposition (\ref{kform}) and
obviously the cutoff version based on truncating this decomposition to a finite sum 
 does depend on the choice of these functions.   Kahane used his inequality to prove that the large $n$ limit is universal,
independent of the choice of the $g_i(x,x')$.

\subsection{Triviality For $b>1$}\label{triviality}
Having developed the tools, our  first goal is to show that Liouville theory, defined as we have assumed by simple normal ordering of the interaction, is trivial for $b>1$.  More precisely,
with $M_{f,\epsilon}(A)=\int_A\d^2x f(x)\normord{e^{2bX_\epsilon(x)}}$, we want to show that for any non-negative $f$ and $A$,  and for $p$ slightly less than $1$,
$\lim_{\epsilon\to 0}\E\left[M_{f,\epsilon}(A)^p\right]=0$.   {\it A priori}, since $M_{f,\epsilon}(A)\geq 0$ and $\E[M_{f,\epsilon}(A)]= \int_A\d^2 x\, f(x)$, we have (with the help of eqn.~(\ref{fracmom}))   
$0\leq \E[M_{f,\epsilon}(A)^p]\leq 1+ \int_A\d^2 x\, f(x)$ for $0\leq p\leq 1$.   Moreover, $\E[M_{f,\epsilon}(A)^p]$  decreases as $\epsilon\to 0$ in view of eqn.~(\ref{momentineq2}).   Hence it
has a limit for $\epsilon\to 0$ that is either positive or zero.  We will show that the limit is $0$.  

The following  proof that the limit vanishes is a rigorous counterpart of the  explanation in section \ref{breakdown}.  As before, it is enough to consider the case that $f=1$ and $A=[0,1]^2$.
   We again pick an integer $q>1$ and
   decompose $A$ as the union of $q^2$ squares $A_i$ that are all translates of $[0,1/q]^2$.   Since $\E[M_{f,\epsilon}(A_i)^p]$ is certainly independent of $i$,
we have for $p<1$
\begin{align}\notag\label{subadd} \E\left[\left( \int_{[0,1]^2} \d^2x \normord{e^{2b X_\epsilon(x)}}\right)^p\right]&\leq \sum_i \E\left[\left( \int_{A_i} \d^2x \normord{e^{2b X_\epsilon(x)}}\right)^p\right]\\ &=q^2
\E\left[\left( \int_{[0,1/q]^2} \d^2x \normord{e^{2b X_\epsilon(x)}}\right)^p\right]. \end{align}
Setting $x=y/q$ and using (\ref{murno}), we get
\begin{align}\notag \label{momcomp}\E\left[\left( \int_{[0,1]^2} \d^2x \normord{e^{2b X_\epsilon(x)}}\right)^p\right]&\leq q^{2-2p}\E\left[\left( \int_{[0,1]^2} \d^2y 
\normord{e^{2b X_{\epsilon}(y/q)}}\right)^p\right]
\\ & =q^{2-2p}\E\left[\left( \int_{[0,1]^2} \d^2y \normord{e^{2b X_{q \epsilon}(y)+2b\Omega_q}}\right)^p\right].\end{align}
Explicitly, $\normord{e^{2b X_{q \epsilon}(y)+2b\Omega_q}}    = e^{2b\Omega_q-2b^2\E[\Omega_q^2] } \normord{e^{2b X_{q \epsilon}(y)}}$, so
\be\label{factoring} \left(\int_{[0,1]^2}\d^2y \normord{e^{2b X_{q \epsilon}(y)+2b\Omega_q}}\right)^p=
e^{2bp\Omega_q-2 b^2p \E[\Omega_q^2]}\left(\int_{[0,1]^2}\d^2y \normord{e^{2b X_{q \epsilon}(y)}}\right)^p ,\ee and
therefore, as $\E[\Omega_q^2]=\log q$, eqn.~(\ref{momcomp}) becomes
\be\label{momcompo}\E\left[\left( \int_{[0,1]^2} \d^2x \normord{e^{2b X_\epsilon(x)}}\right)^p\right]\leq q^{2-2p +2b^2p^2-2b^2p}\E\left[\left( \int_{[0,1]^2} \d^2x \normord{e^{2b X_{q\epsilon}(x)}}\right)^p\right]  \ee
The prefactor $q^{2-2p +2b^2p^2-2b^2p}=q^{2(p-1)(b^2p-1)}$ is familiar from eqn. (\ref{answer}).   As $q>1$, for $b>1$ one can pick $p<1$ so that $q^{2(p-1)(b^2p-1)}<1$.   Having done
so, take $\epsilon\to 0$ in eqn.~(\ref{momcompo}) keeping $q$ fixed.     If the left hand side were to approach a positive limit as $\epsilon\to 0$, the  expectation on the right hand side  would approach the same limit, giving a contradiction.  So the left hand side vanishes as $\epsilon\to 0$, as claimed.

\subsection{Moments}\label{moments}

Henceforth, we assume that $b<1$, and we investigate the behavior of the moments $\E[M_{f,\epsilon}(A)^p]$ as $\epsilon\to 0$.   In this section, we assume that the function $f$ is
smooth and bounded; the singular case is the subject of section \ref{singular}.   

Our first goal is to show, by a rigorous counterpart of the discussion in section \ref{breakdown}, that the limit is $+\infty$ if $p>1/b^2$.   Since $b<1$, $p>1/b^2$ implies in particular
that $p>1$.  So
 $\E[M_{f,\epsilon}(A)^p]$ increases as $\epsilon$ becomes smaller, according to eqn.~(\ref{momentineq}).  Therefore for $\epsilon\to 0$, it
 converges either to a positive real number or to $+\infty$.   We will show that the limit is $+\infty$ for $p>1/b^2$ by a rigorous counterpart of the argument in 
 section \ref{breakdown}.
 
 As before, since $M_{f,\epsilon}(A)$ becomes smaller if $A$ is replaced by a smaller region or $f$ by a smaller positive function, there is no loss in assuming that $f=1$ and
 that $A$ is a small square, which we may as well take to be the unit square $[0,1]^2$.   Again we decompose $A$ as the union of $q^2$ smaller squares all of size $1/q$.
 The first inequality in eqn. (\ref{subadd}) now goes in the opposite direction, since now $p>1$, so following the same steps now leads us to
 \begin{align}\notag \label{momcomp2}\E\left[\left( \int_{[0,1]^2} \d^2x \normord{e^{2b X_\epsilon(x)}}\right)^p\right]&\geq q^{2-2p}\E\left[\left( \int_{[0,1]^2} \d^2y \normord{e^{2b X_{q \epsilon}(y)+2b\Omega_q}}\right)^p\right]\\ &= q^{2(p-1)(b^2p-1)}\E\left[\left( \int_{[0,1]^2} \d^2x \normord{e^{2b X_{q\epsilon}(x)}}\right)^p\right] .\end{align}
Since $q>1$, if $p>1/b^2$, we have $q^{2(p-1)(b^2p-1)}>1$.   Taking $\epsilon\to 0$ with fixed $q$, the inequality implies that the left hand side diverges for $\epsilon\to 0$, as claimed.

The next goal is to show that, with $b<1$, the moments do converge for $p<1/b^2$.  If $p\leq 1$, eqn.~(\ref{momentineq}) shows that the $p^{th}$ moment of $M_{f,\epsilon}(A)$
becomes smaller as $\epsilon$ decreases and therefore converges to a positive limit or zero.   The fact that the limit is positive follows from the proof of $L^1$ convergence
that we give in section \ref{construction}.   Here we consider the case $p>1$, so that $\E[M_{f,\epsilon}(A)^p]$ increases as $\epsilon$ becomes smaller.  We want to show
that for $p<1/b^2$, this moment approaches a finite limit rather than tending to infinity.   The necessary argument, which is inspired by an argument by Bacry and Muzy \cite{BaMu},
 is more elaborate than we have seen so far.

By replacing $A$ with a possibly larger region and $f$ with a possibly larger constant function, we can assume that $f$ is a constant, which may as well be 1, and that
$A$ is a square, which may as well be the unit square $[0,1]^2$.
One preliminary point is that if $p$ is an integer, the statement is true by direct calculation;  the moment  $\lim_{\epsilon\to 0} \E[M_\epsilon(A)^n]$ of integer order $n$  is
given by the integral in eqn.~(\ref{nfold}), which converges if $n<1/b^2$.  
Part of the proof will involve comparing what happens at $p$ to what happens at an integer just less than $p$.   So if $p$ is not an integer, we let $n$ be the  integer that satisfies
\be\label{nearinteger}n-1< p<n ,\ee
and the proof will make use of the fact that the $n-1^{th}$ moment of $M_\epsilon(A)$ converges as $\epsilon\to 0$.   

The first step is again to decompose $A$ as the union of $q^2$ squares $A_i$ each of size $1/q$, now taking $q$ to be an even integer $2r$.  Let $\P$ be the set
of $4r^2$ small squares.   Partition $\P$ into four disjoint subsets $\P_a$, $1\leq i\leq 4$, such that within each $\P_a$, the separation between any pair of small squares is positive.   
This can be done in 
such a way that each $\P_a$ is a translate of $\P_1$.    Define
\be\label{partialsum}S_{a,\epsilon}=\sum_{A_i\in \P_a}M_\epsilon(A_i), \ee
so that $M(A)=S_1+S_2+S_3+S_4$.    Then
\be\label{relone} \E[M_\epsilon(A)^p] \leq 4^{p-1} \sum_{a=1}^4 \E[S_a^p]=4^p \E[S_{1,\epsilon}^p],\ee
where in the last step we use the fact that all $S_a$ have the same distribution.
Now with $n$ as in eqn.~(\ref{nearinteger}), we have
\be\label{elone}S_{1,\epsilon}^p=\left(\sum_{A_i\in \P_1}M_\epsilon(A_i)\right)^p=\left[\left(\sum_{A_i\in \P_1} 
M_\epsilon(A_i)\right)^{p/n}\right]^n\leq \left(\sum_{A_i\in \P_1}M_\epsilon(A_i)^{p/n}\right)^n,\ee where in the last step we use that $p/n<1$.
Expanding this out, we get
\be\label{expanding} S_{1,\epsilon}^p\leq \sum_{A_1,\ldots, A_{n} \in \P_1} M_\epsilon(A_1)^{p/n}M_\epsilon(A_2)^{p/n}\cdots M_\epsilon(A_{n})^{p/n}. \ee
Decompose the right hand side as $U+V$, where $U$ is the sum of all terms in which the $A_i$ are all the same, and $V$ is the sum of all other terms.
As there are $r^2$ little squares in $\P_1$ and $M_\epsilon(A_i)$ is equally distributed for each one, we have  
\be\label{superex}
\begin{split}\E[U]&=r^2\E\left[\left(\int_{[0,1/2r]^2}\d^2x \normord{e^{2b X_\epsilon(x)}}\right)^p\right] \\
&=\frac{1}{4}(2r)^{2(p-1)(b^2 p-1)}
\E\left[\left(\int_{[0,1]^2}\d^2 y \normord{e^{2b X_{2r\epsilon}(y)}}\right)^p\right].
\end{split}
\ee
Here we have followed the same steps as before, setting $y=2rx$, using eqn. (\ref{murno}), and using $\E[\normord{e^{2b \Omega_{2r}}}]=(2r)^{2bp^2-2bp}$.
Because $p>1$, Kahane's inequality (\ref{momentineq})  tells us that the right hand side of eqn.~(\ref{superex}) increases if we replace $X_{2r\epsilon}(y)$ by $X_\epsilon(y)$, and therefore
\be\label{superex2}     
\begin{split} 
\E[U]&\leq \frac{1}{4}(2r)^{2(p-1)(b^2 p-1)}\E\left[\left(\int_{[0,1]^2}\d^2 y \normord{e^{2b X_{\epsilon}(y)}}\right)^p\right]\\
&=\frac{1}{4}(2r)^{2(p-1)(b^2 p-1)}\E[M_\epsilon(A)^p].
\end{split}
\ee
For $p\in (1,1/b^2)$, the exponent in $(2r)^{2(p-1)(b p^2-1)}$ is negative, so we can choose a large $r$ such that 
$C(b,p,r)= 4^{p-1}(2r)^{2(p-1)(b^2 p-1)}$ is less than 1.   As for the remainder $V$ on the right hand side of  eqn.~(\ref{expanding}), since $p/n<1$, we have
$M_\epsilon(A_i)^{p/n}\le 1+M_\epsilon(A_i)$ for all $i$, and therefore
\be\label{restsum}V\leq  {\sum}'_{A_1,\ldots, A_{n}\in\P_1}\prod_{i=1}^n\left(1+ M_\epsilon(A_i)\right),\ee
where the symbol $\sum'$ represents a sum restricted to terms in which the $A_i$ are not all the same.  For each choice of $A_1,A_2,\ldots, A_n$, the corresponding contribution to the
right hand side of eqn.~(\ref{restsum}) is the sum of $2^n$ terms, each of which
is the product of at most $n$ factors of $M_\epsilon(A_i)$ for various different
 $A_i\in\P_1$.    Considering any one such term, let $A'_1,\ldots, A'_{n'}$, with $n'\leq n$, be the little squares that
appear in that term. The contribution of that term to $\E[V]$ can be described as  an explicit integral as in
eqn.~(\ref{nfold}):
\be\label{estsum0}\E\left[\prod_{i=1}^{n'} M_\epsilon (A'_i)\right]=\int_{A'_1}\cdots\int_{A'_{n'}} \d^2 x_1\cdots \d^2 x_{n'} 
\prod_{1\leq i<j\leq n'}\prod_{i<j}\exp(4b^2\E[X_\epsilon(x_i) X_\epsilon(x_j)]   ).\ee  
We would like to have an upper bound on this integral that is independent of $\epsilon$.   Since $\E[X_\epsilon(x_i) X_\epsilon(x_j)]  $ increases
as $\epsilon$ becomes smaller, we can get an upper bound by simply setting $\epsilon=0$, whereupon $\E[X_\epsilon(x_i) X_\epsilon(x_j)] $ reduces to
$\log\frac{1}{|x_i-x_j|}$ and eqn.~(\ref{estsum0}) becomes
\be\label{estsum}\E\left[\prod_{i=1}^{n'} M_\epsilon (A'_i)\right]\leq \int_{A_1'}\cdots \int_{A'_{n'}} \d^2 x_1\cdots \d^2 x_{n'} \prod_{1\leq i< j\leq n'} \frac{1}{|x_i-x_j|^{4b^2}}.
\ee
As a consequence of the remark following eqn.~(\ref{nfold2}), the integral on the right hand side converges, because at most $n-1<1/b^2$ of the points $x_i$ are contained in any one of the $A'_i$, and the distances between the different $A'_i$ are positive. 
(It can happen that the $A_i'$ are all the same, but this is only possible if $n'\leq n-1$.) 
Summing many such terms, $V$ is then bounded above by a positive constant $D(b,p,r)$ that depends on $b , p,$ and $r$ but not on $\epsilon$.

Finally, then, for $p\in (1,1/b^2)$ we get an inequality of the form 
\be\label{finelineq} \E[M_\epsilon(A)^p]\leq C(b,p,r)\E[M_\epsilon(A)^p]+D(b,p,r),\ee
where $C(b,p,r)<1$ and both $C(b,p,r)$ and $D(b,p,r)$ depend only on $b,p$, and $r$ and not $\epsilon$.     This implies an upper bound
\be\label{finalresult} \lim_{\epsilon\to 0} \E[M_\epsilon(A)^p]\leq \frac{D(b,p,r)}{1-C(b,p,r)},\ee
showing that the left hand side  remains bounded as $\epsilon\to 0$.     

What happens for $p<1$?  
For $0< p\leq 1$, the quantity $\E[M_\epsilon(A)^p]$ is bounded above by $1+\E[M_\alpha(A)]=2$ by virtue of eqn.~(\ref{fracmom}), 
and decreases as $\epsilon$ becomes smaller because of Kahane's inequality (\ref{momentineq2}). So its limit as $\epsilon\to 0$
certainly exists.   In this range of $p$, the nontrivial question is whether the limit vanishes, as we know happens for $b>1$. To show that this limit is 
nonvanishing for all $b<1$ requires a fairly difficult argument
that is explained in section \ref{construction}.

\subsection{Convergent Integrals of Singular Functions}\label{singular}
In this section, we begin the analysis of the integral
\be\label{alphaintegral}
I(\upalpha) = \int_{|x|\le 1}  \d^2x\, |x|^{-2b\upalpha} \normord{e^{2bX(x)}},
\ee
with $\upalpha >0$. Note that this integral is always positive, but may potentially take the value $+\infty$. We will  show that  for $\upalpha \in (b, Q)$ (where we recall that  $Q = b + 1/b$), this does not happen, and in fact, for such $\upalpha$, 
\be\label{ialpha}
\E[I(\upalpha)^p] <\infty \text{ for all $p\in (0,(Q-\upalpha)/b)$.}
\ee   This was first shown in \cite{KRV}.
Before attempting a proof, we will first discuss some qualitative aspects of this statement.  First of all, since the upper bound on $\upalpha$ in this statement is
$\upalpha<b+1/b$, in particular it is possible to have $\upalpha\geq 1/b$.  For such values of $\upalpha$, the $|x|^{-2b\upalpha}$ singularity  is not integrable in the usual
sense:  $\int_{|x|\leq 1}\d^2x \frac{1}{|x|^{2b\upalpha}}=\infty$.   The claim in eqn. (\ref{ialpha}) only makes sense because including the factor $\normord{e^{2b X(x)}}$ improves
the situation.   

This is only true, however, if we interpret $I(\upalpha)$ correctly.   A naive imitation of what we have done up to this point would be to define
\be\label{wrongidea} I(\upalpha)\overset{?}{=} \lim_{\epsilon\to 0}  \int_{|x|\leq 1}  \d^2 x\, |x|^{-2b\upalpha}\normord{e^{2bX_\epsilon(x)}}, \ee
but this is in fact not correct.   Indeed, the right hand side of eqn.~(\ref{wrongidea}) is divergent if $\upalpha\geq 1/b$.   Once we replace $X$ by $X_\epsilon$, the random variable
$\normord{e^{2b X_\epsilon(x)}}$ is
typically a continuous and nonzero function near $x=0$ and does not help with the convergence of the integral.

A simple and correct idea is to introduce a spatial cutoff $\eta$, restrict the integral defining $I(\upalpha)$ to $\eta>0$, and then take the limit $\eta\to 0$.  Thus
\be\label{betteridea} I(\upalpha) =\lim_{\eta\to 0}\lim_{\epsilon\to 0} \int_{\eta\leq |x|\leq 1}\d^2 x |x|^{-2b\upalpha} \normord{e^{2b X_\epsilon(x)}}. \ee
It is important that the limit $\epsilon\to 0$ is taken before the limit $\eta\to 0$.   Then the moment that we want to study is 
\begin{align} \E[I(\upalpha)^p]&=   \E\left[\lim_{\eta\to 0}\lim_{\epsilon\to 0} \left(  \int_{\eta\leq |x|\leq 1}\d^2 x |x|^{-2b\upalpha} \normord{e^{2b X_\epsilon(x)}}  \right)^p\right]    \cr &\leq \lim_{\eta\to 0}\lim_{\epsilon\to 0} \E\left[\left(  \int_{\eta\leq |x|\leq 1}\d^2 x |x|^{-2b\upalpha} \normord{e^{2b X_\epsilon(x)}}  \right)^p\right] .\label{zalpha}\end{align}
where the inequality at the end is Fatou's inequality\footnote{In general, the statement of Fatou's inequality involves $\lim\inf$ on both sides, but here
that is not necessary as 
the limits exist on both sides. The limit  as $\epsilon\to 0$ of the expectation in the second line exists because this expectation is monotonically increasing or decreasing (depending on $p$) for $\epsilon\to 0$, by virtue of Kahane's inequality, and then this expectation monotonically increases as $\eta\to 0$ since the integration region is becoming larger.  The limit inside the expectation in the first line exists because of the convergence of the integral to a limit as $\epsilon \to 0$ and then by monotonicity as $\eta \to 0$. 
 In section \ref{martingale}, we have
seen examples in which Fatou's inequality is a strict inequality, namely martingales $\Phi_n$ such that $\E[\Phi_n]=1$ for all $n$ but $\lim_{n\to\infty}\Phi_n=0$.
Thus $0=\E[\lim_{n\to\infty}\Phi_n]<\lim_{n\to\infty}\E[\Phi_n]=1$. However, in eqn.~(\ref{zalpha}), it can be shown that the inequality is actually satisfied as an equality as long as $p$ is in the
range that makes the right hand side finite. But we will not need to show that; the inequality is sufficient for our purposes.} from measure theory. 

However, we should discuss the validity of eqn. (\ref{betteridea}).
   In section \ref{construction}, we will explain how to prove that the cutoff random measure $\normord{e^{2b X_\epsilon(x)}}\d^2x$ converges
as $\epsilon\to 0$ to a Liouville measure $M(x)\d^2x=\normord{e^{2b X(x)}}\d^2x$ that is valued in positive random variables.  Given this measure,
 the integral $\int \d^2 x M(x) f(x)$ makes sense for any positive function $f$ that is measurable in the usual sense, and defines a positive random variable, though in general
 this random variable may have a positive probability (or even probability $1$) to equal $+\infty$.  In the present case, we want to carry out this construction with the positive
 measurable function $f(x)$ that equals $1/|x|^{2b\upalpha}$ if $|x|\leq 1$ and vanishes otherwise.  
 
 The monotone convergence theorem of measure theory says that for any measure -- in our case the Liouville measure $\d^2 x\,M(x)$ -- if $f_1, f_2,\ldots $ is an increasing sequence
 of nonnegative measurable functions that converges pointwise to $f$, then $\lim_{n\to\infty}\int \d^2 x\,M(x) f_n(x)=\int\d^2 x \,M(x) f(x)$.    The statement that eqn.~(\ref{betteridea}) is
 a valid way to define the random variable $I(\upalpha)$ is an example of this idea with $f(x)$ as defined in the last paragraph and with $f_\eta(x)$ being the function that equals $f(x)$ for
 $|x|\geq \eta$ and vanishes for $|x|<\eta$.   The functions $f_\eta(x)$ are increasing as $\eta\to 0$ and converge pointwise to $f(x)$.   In eqn.~(\ref{betteridea}),
 the limit $\epsilon\to 0$ produces $\int \d^2 x M(x) f_\eta(x)$, and then the limit as $\eta\to 0$ produces $\lim_{\eta\to 0}\int \d^2x M(x) f_\eta(x)=\int\d^2 x M(x) f(x)=I(\upalpha)$.
 
 We note that for $\upalpha\geq 1/b$, the expectation value of $I(\upalpha)$ is $+\infty$:
 \be\label{expialpha} \E[I(\upalpha)]=\E\left[\int_{|x|\leq 1}\d^2 x \normord{e^{2bX(x)}} \frac{1}{|x|^{2b\upalpha}}\right]=\int_{|x|\leq 1}\d^2x \frac{1}{|x|^{2b\upalpha}}= +\infty.\ee
For any $\alpha\in (b,1/b+b)$, eqn.~(\ref{ialpha}) asserts that
there is a range of $p>0$ with $\E[I(\upalpha)^p]<\infty$, but if $\upalpha\geq 1/b$, this range is limited to a subset of $p\in (0,1)$.   The existence of any $p>0$ with
$\E[I(\upalpha)^p]<\infty$ implies that the probability for $I(\upalpha)$ to equal $+\infty$ is $0$.   Thus for $\upalpha\in [1/b,1/b+b)$, $I(\upalpha)$ is a random variable that is
finite with probability $1$, but whose expectation value is infinite.

We should also comment on the lower bound $\upalpha>b$ in the statement (\ref{ialpha}). For any $p\in (1,1/b^2)$, there is $\upalpha>b$ with $p<(Q-\upalpha)/b$.   So for
any such $p$, the statement (\ref{ialpha}) implies that there is $\upalpha>b$ with $\E[I(\upalpha)^p]<\infty$.   Making $\upalpha$ smaller makes $1/|x|^{2b\upalpha}$ smaller, and therefore makes $I(\upalpha)$ smaller.
So given that $\E[I(\upalpha)^p]<\infty$ for some $\upalpha>b$, it actually follows that the same is true as well  for all $\upalpha\leq b$ (including the possibility $\upalpha<0$).   
Making $p$ smaller also improves the convergence, as long as $p$ remains positive.   So actually for $\upalpha\leq b$, $\E[I(\upalpha)^p]<\infty$ for all $p\in (0,1/b^2)$.
The same remains true if $p<0$.   We will not prove that last assertion,  but a heuristic explanation was given in section \ref{ltwo}.

Finally we turn to a proof of the assertion (\ref{ialpha}).   First we consider the case $0<p<1$, which as we have already explained encompasses the most delicate situation
with $\upalpha\geq 1/b$.   We use the formula (\ref{zalpha}) for $\E[I(\upalpha)^p]$, whose finiteness we wish to prove.  In this formula, we take $\eta$ to be of the form $2^{-n}$ with
$n=1,2,3,\ldots$, and we consider the limit $n\to \infty$.   Thus for $k=1, 2,3,\ldots$ we define
\be\label{cutoffn}I_{k,\epsilon}(\upalpha) =\int_{2^{-k}\leq |x|\leq 2^{-k+1}}\d^2 x \,|x|^{-2b \upalpha} \normord{e^{2b X_\epsilon(x)}}, \ee
in terms of which
\be\label{cutoffintegral} I(\upalpha)=\lim_{n\to\infty}\lim_{\epsilon\to 0} \sum_{k=1}^n I_{k,\epsilon}(\upalpha), \ee
and so  by Fatou's inequality,
\be\label{momentcut} \E[I(\upalpha)^p]\leq \lim_{n\to\infty}\lim_{\epsilon\to 0} \E\left[\left(\sum_{k=1}^n I_{k,\epsilon}(\upalpha)\right)^p\right]\leq \lim_{n\to\infty}\lim_{\epsilon\to 0}\sum_{k=1}^n\E\left[
I_{k,\epsilon}(\upalpha)^p\right],\ee
where the last step is valid for $0<p<1$.   

By familiar reasoning, we have
\begin{align}\notag &\E\left[\left( \int_{\frac{c}{2}\leq |x|\leq c}\d^2x |x|^{-2b\upalpha}\normord{e^{2b X_\epsilon(x)}} \right)^p\right]\notag \\
&=c^{(2-2b\upalpha)p} 
\E\left[\left( \int_{\frac{1}{2}\leq |y|\leq 1}\d^2 y |y|^{-2b\upalpha}\normord{e^{2b X_\epsilon( cy)}} \right)^p\right] \notag \\ \notag &
=c^{(2-2b\upalpha)p} \E\left[\left( \int_{\frac{1}{2}\leq |y|\leq 1}\d^2 y |y|^{-2b\upalpha}\normord{e^{2b X_{\epsilon/c} ( y)+2b\Omega_{1/c}}} \right)^p\right]       \\ &
=c^{2p(1-b\upalpha+b^2-b^2p)} \E\left[\left( \int_{\frac{1}{2}\leq |y|\leq 1}\d^2 y |y|^{-2b\upalpha}\normord{e^{2b X_{\epsilon/c} ( y)}} \right)^p\right].   \label{scalec}\end{align}
In the following, we use the fact that for  $c<1$, $c^{2p(1-b\upalpha+b^2-b^2p)}<1$  precisely if $p<(Q-\upalpha)/b$.   

Setting $c=2^{-(k-1)}$, eqn.~(\ref{scalec}) means that in eqn.~(\ref{momentcut}), we can replace $I_{k,\epsilon}(\upalpha)^p$ with $2^{-2p(k-1)(1-b\upalpha+b^2-b^2p)}I_{1,2^{k-1}\epsilon}(\upalpha)$.
Since we will be taking $\epsilon\to 0$ with fixed $n$ and therefore with an upper bound on $k$, it  makes no difference here to replace $I_{1,2^{k-1}\epsilon}(\upalpha)$ with $I_{1,\epsilon}(\upalpha)$.  The sum over $k$  on the right hand side of eqn.~(\ref{momentcut}) 
is then just a geometric series and the $n\to\infty$ limit consists of continuing the geometric series up to infinity.
 So  eqn.~(\ref{momentcut}) becomes an inequality
\be\label{geomsum}\E[I(\upalpha)^p]\leq
 \sum_{k=1}^\infty  2^{-2p(k-1)(1-b\upalpha+b^2-b^2p)}\lim_{\epsilon\to 0} \E\left[ I_{1,\epsilon}(\upalpha) ^p  \right]. \ee
The geometric series converges if $p<(Q-\alpha)/b$, and $\lim_{\epsilon\to 0}\E[I_{1,\epsilon}(\upalpha)^p]$ is also finite,  because $I_{1,\epsilon}(\upalpha)$ is defined
by an integral over the region $|x|>1/2$, away from the singularity.   Indeed,    since we have assumed $p<1$, we have for all $\epsilon$,
 $ \E[I_{1,\epsilon}(\upalpha)^p]<1+  \E[I_{1,\epsilon}(\upalpha)]=1+\int_{\frac{1}{2}\leq |x|\leq 1}\d^2 x |x|^{-2b\upalpha}<\infty$,  though
pending the analysis in section \ref{construction}, we do not yet know that $\lim_{\epsilon\to 0} \E[I_{1,\epsilon}(\upalpha)^p]\not=0.$
  So eqn.~(\ref{geomsum}) implies that $\E[I(\upalpha)^p]<\infty$ in the claimed range of $p$. 

The other case $p>1$ is less delicate as it only arises if $\upalpha<1/b$.  Let $I_{\eta,\epsilon}(\upalpha)$ be $I(\upalpha)$ modified by the $\eta$ and $\epsilon$ cutoffs, and
for $\eta<c<1$, define 
\be\label{twoterms} I_{\eta,\epsilon}(\upalpha)=  J_{\eta,\epsilon,c}(\upalpha)+K_{\epsilon,c}(\upalpha),\ee
with
\be\label{restrictedintegrals} J_{\eta,\epsilon,c}(\upalpha)=\int_{\eta\leq |x|\leq c}\d^2 x |x|^{-2b\upalpha} \normord{e^{2bX_\epsilon(x)}},~~~~~K_{\epsilon,c}(\upalpha)=\int_{c\leq |x|\leq 1}\d^2 x |x|^{-2b\upalpha} \normord{e^{2bX_\epsilon(x)}}.\ee
Hence
\be\label{usefuldecomp}\E[I_{\eta,\epsilon}(\upalpha)^p]=\E[(J_{\eta,\epsilon,c}(\upalpha)+K_{\epsilon,c}(\upalpha))^p]\leq 2^{p-1}\left( \E[J_{\eta,\epsilon,c}(\upalpha)^p]+\E[K_{\epsilon,c}(\upalpha)^p]\right). \ee
The familiar scaling arguments show that 
\be\label{scalapp} \E[J_{\eta,\epsilon,c}(\upalpha)^p]= c^{2p(1-b\upalpha+b^2-b^2p)} \E[I_{\eta/c,\epsilon/c}(\upalpha)^p].\ee   
Replacing $\eta/c$ by $\eta$ and $\epsilon/c$ by $\epsilon$ makes $ \E[I_{\eta/c,\epsilon/c}(\upalpha)^p]$  larger; in the case of $\eta$, this is because the replacement makes
the integration region larger, while in the case of $\epsilon$, it is because of Kahane's inequality (\ref{momentineq}). 
So the inequality (\ref{usefuldecomp}) becomes
\be\label{finaldecomp} \E[I_{\eta,\epsilon}(\upalpha)^p]\leq 2^{p-1}\left( c^{2p(1-b\upalpha+b^2-b^2p)} \E[I_{\eta,\epsilon}(\upalpha)^p] +\E[K_{\epsilon,c}(\upalpha)^p]\right).\ee
On the right hand side, since the condition $p<(Q-\upalpha)/b$ together with $\upalpha>b$ implies that $p<1/b^2$, the result of section \ref{moments} implies that
 $\lim_{\epsilon\to 0} \E[K_{\epsilon,c}(\upalpha)^p]<\infty$. 
Choosing $c$ so that $2^{p-1} c^{2p(1-b\upalpha+b^2-b^2p)} <1$, the inequality   (\ref{finaldecomp}) gives an upper bound on $\E[I_{\eta,\epsilon}(\upalpha)^p]$ that
is independent of the cutoff parameters $\eta$ and $\epsilon$.  So  $\E[I(\upalpha)^p]<\infty$.

\subsection{Divergent Integrals of Singular Functions}\label{divergence}
We will now investigate, somewhat informally, the counterpart of eqn.~(\ref{ialpha}), which says that $I(\upalpha)=\infty$ with probability $1$ if $\upalpha \ge Q$. 
The analysis will also shed light on the convergence of $I(\upalpha)$ for $\upalpha<Q$.

We parametrize the $x_1,x_2$ plane with polar coordinates $r,\theta$ centered at $x=0$ and set $r=e^{-s}$.  
 For $s\ge 0$, let $B_s = X_{e^{-s}}(0)$ be the circle average of the Gaussian free field $X$ on the circle  $r=e^{-s}$.  Also, define  the ``remainder''
\be
Y(s,\theta) = X(e^{-s} e^{i\theta}) - B_s.
\ee
In terms of the Fourier expansion $X(s,\theta)=X_0(s)+\sum_{n\not=0} e^{\i n\theta}X_n(s)$ of eqn.~(\ref{fourier}), $B_s=X_0(s)$ and $Y(s,\theta)=\sum_{n\not=0} e^{\i n \theta}X_n(s)$.
Since $X(s,\theta)=B_s+Y(s,\theta)$ and $B_s$ and $Y(s,\theta)$ are independent, we have $\normord{e^{2b X(s,\theta)}}=\normord{e^{2b B_s}} \,\normord{e^{2b Y(s,\theta)}}$.
Here $\normord{e^{2b B_s}}=e^{2b B_s-2b^2 \E[B_s^2]}$.   We recall from section \ref{martingale} that $B_s$ has the statistics of Brownian motion, normalized so that
$B_0=0$ because of the condition\footnote{If we did not have the condition (\ref{condx}), we would still have $\E[(B_s-B_0)^2]=s$  because of the normalization
of the action.   We would also have $\E[(B_s-B_0)B_0]=0$, because in Brownian motion the increment $B_s-B_0$ is independent of $B_0$.
So we would get $\E[B_s^2]=s+\E[B_0^2]$.   A constant factor $e^{2b^2\E[B_0^2]}$ would then appear in the following formulas; of course, this would not affect
the discussion of convergence of integrals.}
(\ref{condx}) and with $\E[B_s^2]=s$ because of the normalization of the action (\ref{zeroform}). 
  So
\begin{align}\notag I(\upalpha)=\int_{|x|<1}\d^2 x |x|^{-2b\upalpha} \normord{e^{2b X(x)}} &=\int_0^\infty \d s e^{-s(2-2b\upalpha +2b^2 s) } e^{2b B_s} \int_0^{2\pi}\d\theta \normord{e^{2b Y(s,\theta)}} e^{2b^2 \E[B_0^2]}\\ &
= \int_0^\infty \d s e^{-2s(Q-\upalpha) } e^{2b B_s} \int_0^{2\pi}\d\theta \normord{e^{2b Y(s,\theta)}}  . \label{intform}\end{align}
The integral over $s$ diverges exponentially if $\upalpha>Q$.  
Indeed, the  random variable $B_s$ is typically of order $s^{1/2}$, too small to affect the convergence of the integral.
Since the zero-mode of $X$ has been removed in defining the random variable $Z_s=\int_0^{2\pi}\d\theta \normord{e^{2b Y(s,\theta)}}$, this random variable has a stationary distribution,
invariant under constant shifts of $s$,  reflecting the scale-invariance of Liouville theory.  So on the average $Z_s$ is not growing or decaying as $s$ becomes large and does not help with the convergence of the integral. 

  If $\upalpha=Q$, the integral does not diverge exponentially.   But it does still diverge with probability 1,
because $Z_s$ has only short range correlations and is independent of $B_s$.   So if we partition the half-line $s\geq 0$ as the union of infinitely many closed intervals
$[n,n+1]$, then with probability 1, in infinitely many of those intervals one has $Z_s>\delta$ (for some chosen $\delta$) and $B_s>0$, ensuring a divergence of the integral.

For $\upalpha<Q$, the integral in eqn.~(\ref{intform}) converges with probability $1$, so $I(\upalpha)<\infty$ with probability $1$.  This was already clear from section \ref{singular},
where we showed that if $\upalpha<Q$, then  $\E[I(\upalpha)^p]<\infty$ for some range of $p>0$.

\subsection{Construction of Liouville Measure for all $b<1$}\label{construction}
Finally, here we will sketch  the proof of the convergence of the random variable  $M_{f,\epsilon}(A)=\int_A \d^2x f(x)\normord{e^{2b X_\epsilon(x)}} $ to a nonzero limit for any $0\le b< 1$ and assuming that $f$ is smooth with nonzero integral over $A$.  This is a much stronger claim than the statements about moments that have been discussed so far. The proof we give here is inspired by and very similar to an argument of Berestycki~\cite{berestycki17}. 
 In contrast to section \ref{ltwo}, which was based on convergence in the $L^2$ norm, here we will consider convergence in the $L^1$ norm, defined for any real-valued random variable
 $\Phi$ as the expectation of the absolute value of $\Phi$.   Random variables are complete in the $L^1$ norm just as in the $L^2$ norm.
 So if
\be\label{dlim1}
\lim_{\epsilon, \epsilon'\to \infty} \E[|M_{f,\epsilon}(A)-M_{f,\epsilon'}(A)|] = 0,
\ee
which shows that $M_{f,\epsilon}(A)$ is a Cauchy sequence in the $L^1$ norm, then as $\epsilon\to 0$, the family of random variables  $M_{f,\epsilon}(A)$ will converge in $L^1$
 to a limit $M_f(A)$. Convergence in $L^1$ guarantees in particular that $\E[M_{f,\epsilon}(A)]$ converges to $\E[M_f(A)]$. But
\be
 \E[M_{f,\epsilon}(A)] = \int_A \d^2x f(x)
\ee
for any $\epsilon$, and therefore $\E[M_f(A)]\ne 0$. In particular, $M_f(A)$ has a nonzero chance of being nonzero.   This compares to what happens for $b>1$,
where $\lim_{\epsilon\to 0} M_{f,\epsilon}(A)=0$, as found in sections  \ref{breakdown} and \ref{triviality}.

To prove eqn.~(\ref{dlim1}), fix  $0<\epsilon'/2<\epsilon < \epsilon'$. Let $I_\epsilon := M_{f,\epsilon}(A)$ and $I_{\epsilon'} := M_{f,\epsilon'}(A)$. We will define two related quantities $J_\epsilon$ and $J_{\epsilon'}$, and show that if $\epsilon$ and $\epsilon'$ are small, then
\begin{itemize}
\item $I_\epsilon$ is close to $J_\epsilon$ in $L^1$,
\item $I_{\epsilon'}$ is close to $J_{\epsilon'}$ in $L^1$, and
\item $J_\epsilon$ is close to $J_{\epsilon'}$ in $L^2$.
\end{itemize}
For any random variable, the $L^1$ norm is smaller than or equal to the $L^2$ norm. Thus, if we are able to show the above, then by the triangle inequality for the $L^1$ norm, $I_\epsilon$ and $I_{\epsilon'}$ are close in the $L^1$ norm, proving eqn.~(\ref{dlim1}). 

For each $x\in  A$, we define a random variable $G(x)$, which is $1$ if $X_a(x) \le \zzeta \log (1/a)$ for all $a\in [\epsilon', 3\epsilon']$, and $0$ otherwise, where we will choose the number $\zzeta \in ( 2b, 4b)$ later.  We define
\be
J_\epsilon = \int_A \d^2x \normord{e^{2b X_\epsilon(x)}} G(x), \ \  J_{\epsilon'} =  \int_A \d^2x \normord{e^{2b X_{\epsilon'}(x)}} G(x).
\ee
In particular, note that $J_\epsilon \le I_\epsilon$ and $J_{\epsilon'}\le I_{\epsilon'}$. Thus,
\be
\E|I_\epsilon-J_\epsilon| = \E[I_\epsilon - J_\epsilon] = \int_A \d^2x f(x) \E[ \normord{e^{2b X_\epsilon(x)}} (1-G(x))]. 
\ee
We will use Girsanov's theorem to evaluate the expectation on the right. The simplest version of Girsanov's theorem says the following. Suppose that $B_t$ is a standard Brownian motion 
and $\tilde{B}_t = B_t + \nu t$ is Brownian motion with drift $\nu$. Let $F$ be a function of the whole path $(B_t: 0\le t\le T)$ for some $T$. Let $\tilde{F}$ be the same function of $(\tilde{B}_t:0\le t\le T)$. Then\footnote{The assertion that $B_t$ is standard Brownian motion means that $B_t$ is Brownian motion normalized so that $B_0=0$ and $\E[B_t^2]=t$. Thus  $B_t$ is governed
by the distribution function $\frac{1}{Z}\exp\left(-\frac{1}{2}\int_0^T\d t \left(\frac{\d B_t}{\d t}\right)^2\right)$, where $Z$ is a normalization constant, and
$\t B_t=B_t+\nu t$ is governed by the distribution function $\frac{1}{Z}\exp\left(-\frac{1}{2}\int_0^T\d t \left(\frac{\d \tilde B_t}{\d t}-\nu\right)^2\right)$.  Hence
$\E[F(\tilde B_t)]=\frac{1}{Z}\int D\tilde B_t \,F(\tilde B_t)\exp\left(-\frac{1}{2}\int_0^T\d t \left(\frac{\d \tilde B_t}{\d t}-\nu\right)^2\right)=\frac{1}{Z}\int D\tilde B_t \,F(\tilde B_t)e^{\nu \tilde B_t}e^{-\frac{1}{2}\nu^2 T} \exp\left(-\frac{1}{2}\int_0^T
\d t \left(\frac{\d \tilde B_t}{\d t}\right)^2\right)$.   Renaming the integration variable $\t B_t$ as $B_t$ and using $T=\E[B_T^2]$,
we conclude that $\E[F(\t B_t)]=\E[e^{\nu B_T - \frac{1}{2}\nu^2 \E[B_T^2]} F(B_T)]$, as claimed in the text, where  the expectation on the right of the final statement
 is taken with respect to  standard Brownian motion. }
\be
\E[\tilde{F}] = \E[e^{\nu B_T - \frac{1}{2}\nu^2 \E[B_T^2]} F].
\ee
Fixing any $x\in A$, let $B_t = X_{e^{-t}}(x)$, and let $\tilde{B}_t = B_t + 2b t$. Recall that $G(x)=1$ if $B_t \le \zzeta t$ for all $t \in [\log (1/3\epsilon'), \log (1/\epsilon')]$, and $0$ otherwise. Analogously, define $\tilde{G}(x) = 1$ if $\tilde{B}_t \le \zzeta t$ for all $t \in [\log (1/3\epsilon'), \log (1/\epsilon')]$ and $0$ otherwise. Let $T = \log (1/\epsilon)$. Then by Girsanov's theorem,
\be
\begin{split}
&\E [\normord{e^{2b X_\epsilon(x)}} (1-G(x))] \\
&= \E[e^{2b B_T - 2b^2 \E[B_T^2]} (1-G(x))]\\
&= \E[1-\tilde{G}(x)] = \mathbb{P}(\tilde{B}_t > \zzeta t \text{ for some } t \in [\log (1/3\epsilon'), \log (1/\epsilon')]),
\end{split}
\ee
 where we used the simple fact that if a random variable $Y$ can only be $0$ or $1$, then $\E[Y] = {\mathbb P}(Y=1)$ (with $\mathbb{P}(E)$ denoting the probability of an event $E$). But since $\zzeta>2b$ and $\tilde{B}_t$ is a Brownian motion with drift $2b$, it is very unlikely to have $\tilde B_t>\zzeta t$ at late $t$, and more specifically it follows from standard facts about Brownian motion that  
\be
\mathbb{P}(\tilde{B}_t > \zzeta t \text{ for some } t \in [\log (1/3\epsilon'), \log (1/\epsilon')]) \le e^{-C(\zzeta, b) \log(1/\epsilon)},
\ee
where $C(\zzeta,b)$ is a positive constant that depends only on $\zzeta$ and $b$. 
 This shows that 
 \be\label{l1bound}
 \E\left[|I_\epsilon - J_\epsilon|\right]  \le  e^{-C(\zzeta, b) \log(1/\epsilon)},
 \ee
 and a similar bound holds for  $\E\left[|I_{\epsilon'} - J_{\epsilon'}|\right]$. Next, note that
\be\label{jepsilon}
\begin{split}
&\E[(J_{\epsilon} - J_{\epsilon'})^2 ]\\
&= \int_{A\times A}\d ^2x \d ^2y f(x)f(y) \\
&\qquad \qquad \qquad \cdot \E[(\normord{e^{2b X_\epsilon(x)}} - \normord{e^{2b X_{\epsilon'}(x)}})(\normord{e^{2b X_\epsilon(y)}} - \normord{e^{2b X_\epsilon'}(y)})G(x)G(y)].
\end{split}
\ee
Take any $x,y\in A$ such that $|x-y|> 2\epsilon'$. Then, given the values of $X$ outside the union of the two open balls of radius $\epsilon'$ centered at $x$ and $y$, the conditional expected value of $\normord{e^{2b X_\epsilon(x)}}$ is $\normord{e^{2b X_{\epsilon'}(x)}}$, and the conditional expected value of $\normord{e^{2b X_\epsilon(y)}}$ is $\normord{e^{2b X_{\epsilon'}(y)}}$. (This follows from the fact that $G(x)$ and $G(y)$ only depend on data outside of the two balls of radius $\epsilon'$, together with the martingale property of the normal ordered
exponential of Brownian motion, which was explained in section \ref{martingale}.)
 Therefore, for such $x$ and $y$, the integrand is zero. 

Now, take any $x,y\in A$ such that $|x-y|\le 2\epsilon'$. Let $B_t = X_{e^{-t}}(x)$ and $T_0 = \log (1/3\epsilon')$. Note that $G(x)G(y) \le H$, where $H$ is the random  variable that is $1$ if $B_{T_0}\le \zzeta T_0$ and $0$ otherwise. Now, let us open up the parentheses inside the expectation on the right in eqn.~(\ref{jepsilon}). Consider the first term:
\be\label{step1}
\E\left[\normord{e^{2b X_\epsilon(x)}}\normord{e^{2b X_\epsilon(y)}} G(x)G(y))\right] \le \E[e^{2b (X_\epsilon(x) +X_{\epsilon}(y)] - 4b^2\log(1/\epsilon)} H].
\ee
We will now use the following simple fact about Gaussian random variables. Suppose that $X$ and $Y$ are jointly Gaussian random variables with expected value zero. Take any $\nu$, and let 
\be
a = \nu \E[XY].
\ee
Let $c^2$ be the variance of $Y$. Let $Z$ be another Gaussian random variable, with expected value $a$ and variance $c^2$. Then for any function\footnote{The following
statement can be proved by integrating over $X$ in the joint distribution of $X$ and $Y$, which by hypothesis is Gaussian.}
 $F$,
\be
\frac{\E[e^{\nu X} F(Y)]}{\E[e^{\nu X}]} = \E[F(Z)]. 
\ee
Let us now apply this identity with $X = X_\epsilon (x) + X_\epsilon(y)$, $Y = B_{T_0}$, and $\nu = 2b$.  Let $F$ be the function such that $H = F(B_{T_0})$. Then 
\be\label{step2}
\frac{\E[e^{2b(X_\epsilon(x)+X_{\epsilon}(y))} H]}{\E[e^{2b(X_\epsilon(x)+X_{\epsilon}(y))}]} = \E[F(Z)],
\ee
where $Z$ is a Gaussian random variable with expected value 
\be
a = 2b\E[X_{3\epsilon'}(x)(X_\epsilon(x) + X_\epsilon(y))]
\ee
and variance $c^2 = \mathrm{Var}(X_{3\epsilon'}(x))$. Now, by eqn.~(\ref{usefulfact}) and the fact that the circles of radius $\epsilon$ around $x$ and $y$ are both inside the disk of radius $3\epsilon'$ around $x$, we get that 
\be
\E[X_{3\epsilon'}(x)X_\epsilon(x)] = \E[X_{3\epsilon'}(x)X_\epsilon(y)] = \log \frac{1}{3\epsilon'}, 
\ee
and thus,
\be
a = 4b\log \frac{1}{3\epsilon'}.
\ee
Also by eqn.~(\ref{usefulfact}), $c^2 = \log (1/3\epsilon')$. Thus, $Z$ is a Gaussian random variable with mean $4b \log (1/3\epsilon')$ and variance $\log (1/3\epsilon')$. Consequently, 
if $\zzeta<4b$ and $\epsilon'$ is small, $Z$ is very unlikely to be less than $\zzeta\log(1/3\epsilon')$.  
By standard estimates for the tail of the Gaussian distribution,
\be\label{step3}
\begin{split}
\E[F(Z)] &= \mathbb{P}(Z\le \zzeta \log (1/3\epsilon'))\\
&= \exp\biggl(-\frac{1}{2} (4b -\zzeta)^2 \log \frac{1}{3\epsilon'} + o(\log(1/\epsilon'))\biggr). 
\end{split}
\ee
On the other hand, by (\ref{secondineq}), 
\be\label{step4}
\begin{split}
\E[e^{2b(X_\epsilon(x)+X_{\epsilon}(y))}] &= \exp(2b^2 \E[X_\epsilon(x)^2] + 2b^2 \E[X_\epsilon(y)^2] + 4b^2 \E[X_\epsilon(x)X_\epsilon(y))]\\
&\leq \exp(8b^2 \log(1/\epsilon). 
\end{split}
\ee
Combining eqns.~(\ref{step1}), (\ref{step2}), (\ref{step3}) and (\ref{step4}), we get
\be
\begin{split}
&\E[\normord{e^{2b X_\epsilon(x)}}\,\normord{e^{2b X_\epsilon(y)}} G(x)G(y)]  \\
&\le   \exp\biggl(4b^2 \log \frac{1}{\epsilon}-\frac{1}{2} (4b -\zzeta)^2 \log \frac{1}{3\epsilon'} + o(\log(1/\epsilon'))\biggr)\\
&= \exp\biggl(\biggl(4b^2 -\frac{1}{2} (4b -\zzeta)^2\biggr) \log \frac{1}{\epsilon} + o(\log(1/\epsilon))\biggr).
\end{split}
\ee
Dealing similarly with the other terms in the expansion of the integrand in eqn.~(\ref{jepsilon}), we get 
\be
\begin{split}
&\E[(J_\epsilon - J_{\epsilon'})^2)]\\
&\le  \exp\biggl(\biggl(4b^2 -\frac{1}{2} (4b -\zzeta)^2\biggr) \log \frac{1}{\epsilon} + o(\log(1/\epsilon))\biggr)\mathrm{Vol}(\{(x,y)\in A: |x-y|\le 2\epsilon'\})\\
&=   \exp\biggl(\biggl(-2 + 4b^2 -\frac{1}{2} (4b -\zzeta)^2\biggr) \log \frac{1}{\epsilon} + o(\log(1/\epsilon))\biggr).
\end{split}
\ee
Now, if $b<1$ and we choose $\zzeta = 2b$, then 
\be
-2 + 4b^2 -\frac{1}{2} (4b -\zzeta)^2 = -2 + 4b^2 - 2b^2 = -2 + 2b^2 < 0.
\ee
This shows that if $b< 1$, then we can choose $\zzeta$ slightly bigger than $2b$, such that we still have 
\be
-2 + 4b^2 -\frac{1}{2} (4b -\zzeta)^2 < 0.
\ee
Thus, with such a choice of $\zzeta$, and combining the above with eqn.~(\ref{l1bound}), we get that for any $0<\epsilon'/2< \epsilon<\epsilon'$, 
\be\label{penultimate}
\E|I_\epsilon - I_{\epsilon'}| \le e^{-C(\zzeta, b)\log (1/\epsilon)},
\ee
where $C(\zzeta, b)$ is a positive constant that depends only on $\zzeta$ and $b$. This almost proves eqn.~(\ref{dlim1}), except that we have the restriction that $\epsilon' > \epsilon > \epsilon'/2$. To remove this restriction, take any $0<\epsilon < \epsilon'$, and find $\epsilon_1,\ldots, \epsilon_n$ such that $\epsilon= \epsilon_1 < \epsilon_2<\cdots < \epsilon_n =\epsilon'$, and $\epsilon_{k-1} > \epsilon_k/2$ for each $k$. Repeatedly applying eqn.~(\ref{penultimate}) to each $\E|I_{\epsilon_{k-1}} - I_{\epsilon_k}|$, and observing that $n =O( \log (1/\epsilon'))$, we get eqn.~(\ref{dlim1}).

This completes the proof of eqn.~(\ref{dlim1}) for smooth $f$. A consequence is that $M_{1,\epsilon}(A)$ converges to a nonzero limit as $\epsilon \to 0$, where $1$ denotes the function that is $1$ everywhere. We will denote this limit by $M(A)$. The function $M$ associates a non-negative real number to every Borel set $A$, and it satisfies the axioms for being a measure, in the sense of measure theory. Thus, Liouville theory defines a random measure $M$ on the Borel sets of $\R^2$, which is nontrivial (i.e., nonzero) when $b<1$.

\vskip1cm
 \noindent {\it {Acknowledgements.}}  Research of SC supported in part by NSF grants DMS-2113242 and DMS-2153654.
  Research of EW supported in part by NSF Grant PHY-2207584.
 \bibliographystyle{unsrt}

\end{document}